\documentclass[apj]{emulateapj-rtx4}
\usepackage{amsmath}
\usepackage{natbib}
\usepackage{lscape}

\shorttitle{Debris Disk Features and Exozodiacal Dust}
\shortauthors{Ballering, Rieke, \& G\'asp\'ar}
\slugcomment{Accepted for publication in ApJ}

\newcommand{\NumTargs}{22}
\newcommand{\NumClear}{13}
\newcommand{\NumMarginal}{9}
\newcommand{\NumOneBelt}{13}

\begin{document}

\title{Probing the Terrestrial Regions of Planetary Systems:\\Warm Debris Disks with Emission Features}

\author{Nicholas P. Ballering\altaffilmark{1}, George H. Rieke, Andr\'as G\'asp\'ar}
\affil{Steward Observatory, University of Arizona, 933 North Cherry Avenue, Tucson, AZ 85721, USA}
\altaffiltext{1}{ballerin@email.arizona.edu}

\begin{abstract}
Observations of debris disks allow for the study of planetary systems, even where planets have not been detected. However, debris disks are often only characterized by unresolved infrared excesses that resemble featureless blackbodies, and the location of the emitting dust is uncertain due to a degeneracy with the dust grain properties. Here we characterize the \textit{Spitzer} IRS spectra of $\NumTargs$ debris disks exhibiting 10 micron silicate emission features. Such features arise from small warm dust grains, and their presence can significantly constrain the orbital location of the emitting debris. We find that these features can be explained by the presence of an additional dust component in the terrestrial zones of the planetary systems, i.e. an exozodiacal belt. Aside from possessing exozodiacal dust, these debris disks are not particularly unique; their minimum grain sizes are consistent with the blowout sizes of their systems, and their brightnesses are comparable to those of featureless warm debris disks. These disks are in systems with a range of ages, although the older systems with features are found only around A-type stars. The features in young systems may be signatures of terrestrial planet formation. Analyzing the spectra of unresolved debris disks with emission features may be one of the simplest and most accessible ways to study the terrestrial regions of planetary systems.
\end{abstract}

\keywords{circumstellar matter -- infrared: stars -- interplanetary medium -- zodiacal dust}

\section{INTRODUCTION}
\label{sec:introduction}

Terrestrial regions of planetary systems are not well studied. The majority of known exoplanets were discovered using the radial velocity and transit techniques, which are biased to massive and very short-period planets. While some rocky planets have now been discovered in the terrestrial zone (e.g. Kepler-186f; \citealp{quintana2014}) their frequency and statistical properties can only be estimated by extrapolation from planets nearer their stars (e.g. \citealp{petigura2013}). Direct imaging of exoplanets, on the other hand, is currently limited to massive planets on wide orbits outside of the terrestrial zone.

Observations of debris disks provide an alternative method to study planetary systems. Debris disks are the results of the collisional processing of the solid material left over from planet formation, and their locations may be gravitationally influenced by unseen planets (for a recent review of debris disks, see \citealp{matthews2014}). The disks are often partitioned into concentric components, and it is useful to categorize these components by equilibrium temperature \citep{su2014}. ``Cold" components ($\lesssim$100 K), located in the outer parts of planetary systems, are analogous to the Solar System's Kuiper Belt and may trace the radial limits of planet formation or migration \citep{ballering2013}. ``Warm" components ($\sim$200 K) are analogous to the asteroid belt in the Solar System, and their locations may be set by the water ice line \citep{morales2011}. ``Very hot" components ($\sim$1000 K) trace small refractory dust grains located very near to the star (e.g. \citealp{absil2013}). Between the very hot and warm components lies the ``hot" component ($\sim$300 K), also referred to as the terrestrial zone. Dust in this zone may be analogous to the zodiacal dust in the Solar System, likely brought inward from Jupiter Family comets and the main asteroid belt \citep{nesvorny2010}. Studying the terrestrial regions of planetary systems via observations of exozodiacal dust is the subject of this paper.

Debris in terrestrial zones has been largely inaccessible to observation. The {\it Herschel} and {\it Spitzer} space telescopes have imaged disks at mid- to far-IR wavelengths (e.g., \citealp{su2005}; \citealp{su2008}; \citealp{booth2013}; \citealp{morales2013}), but the resolution of these telescopes is not sufficient to resolve the terrestrial regions of these systems. The Atacama Large Millimeter/sub-millimeter Array (ALMA) has high resolution, but works at sub-mm and radio wavelengths so is primarily sensitive to cold components. Debris disks have also been imaged via the starlight they scatter in the visible and near-infrared by the {\it Hubble Space Telescope} (e.g. \citealp{kalas2005}; \citealp{schneider2009}; \citealp{soummer2014}) and large ground-based telescopes equipped with advanced adaptive optics systems (e.g. \citealp{buenzli2010}; \citealp{currie2012}; \citealp{rodigas2014}), but current scattered light observations are limited to regions outside the terrestrial zone by the high dust/star contrast and small inner working angles required. Interferometric observations in the near-infrared have detected very hot dust around some stars (see the current list of detections in \citealp{vanlieshout2014}), but this material resides nearer to the star than the terrestrial region. A promising method to spatially resolve dust in terrestrial regions is interferometry at $\lambda \sim$ 10 $\micron$, such as the detections of exozodiacal dust around $\eta$ Crv \citep{millangabet2011}. A large program to expand such results is being undertaken with the Large Binocular Telescope Interferometer (LBTI), although it will be limited to relatively nearby stars \citep{hinz2009}. The study of exozodiacal dust presented in this paper is complementary to these interferometric observations.

Debris disk characteristics around many stars have been inferred through their spectral energy distributions (SEDs) that show infrared flux in excess of that from the star's photosphere. SED studies, using data primarily from {\it Spitzer} and {\it Herschel}, have succeeded in discovering and characterizing hundreds of debris disks \citep{morales2011,ballering2013,chen2014}. These studies classify disks by temperature and find that they consist of cold components, warm components, or both. However, these components are usually colder than 300 K, which is the characteristic temperature expected for dust in the terrestrial zone.\footnote{The recent analysis by \citet{chen2014} shows some components reaching higher temperatures.}

Conversion of the apparent temperature of a debris disk to its orbital location is uncertain, as there is a degeneracy between the distance from the star and the optical and physical properties of the dust (minimum grain size, grain size distribution, and grain composition). The degeneracy arises from small grains that are superheated above their equilibrium temperatures because they absorb visible starlight more efficiently than they cool by emitting longer wavelength radiation. \citet{booth2013} resolve several cold debris belts with {\it Herschel} and find that they can be up to 2.5 times further from the star than predicted by blackbody fits to their SEDs. \citet{rodriguez2012} find that the disk orbital radii can be up to 5 times their blackbody radii. The degree of difference between the true location of a debris disk and that derived from the SED-measured dust temperature is likely not uniform for all disks, as the size of the smallest grains in a system is determined by the radiation forces exerted on them by the central star and by the collisional and dynamical processes occurring in the system. These uncertainties make it difficult to determine whether or not warm dust is located in the terrestrial zone. Furthermore, this degeneracy operates such that debris disk components seem to be nearer to their stars than they actually are because grains tend to be warmer than their equilibrium blackbody temperatures. Because most warm components have measured temperatures less than 300 K, this degeneracy makes it unlikely that these components are probing the terrestrial zones.

While the emitted flux density of most debris disks can be well modelled with one or two blackbody functions, a minority of disks show solid-state emission features in their spectra, most prominently at $\sim$10 $\micron$ and $\sim$18 $\micron$ from Si-O stretching and O-Si-O bending vibrations in the silicate material, respectively. These features only arise from warm, small (sub-$\micron$ to few-$\micron$) grains. By using the extra information present in the emission features, the degeneracies in modelling a debris disk SED are broken, and the location of the disk and its grain properties can be more accurately determined.

At least sixteen warm debris disks with prominent spectral features have been studied. These include: HR3927 \citep{chen2006}, $\eta$ Crv \citep{chen2006,lisse2012}, HD113766 \citep{chen2006,lisse2008,olofsson2012,smith2012,olofsson2013}, HD172555 \citep{chen2006,lisse2009,smith2012,johnson2012}, $\eta$ Tel \citep{chen2006,smith2009}, HD69830 \citep{beichman2005,lisse2007,beichman2011,olofsson2012}, BD+20 307 \citep{song2005,weinberger2011,olofsson2012}, HD15407A \citep{melis2010,fujiwara2012,olofsson2012}, HD169666 \citep{moor2009,olofsson2012}, [GBR2007] ID8 \citep{meng2012,olofsson2012}, EF Cha \citep{rhee2007b,currie2011}, HD145263 \citep{honda2004}, HD165014 \citep{fujiwara2010}, HD23514 \citep{meng2012,rhee2008}, HD72905 \citep{beichman2006}, and the very well-studied debris disk around $\beta$ Pic \citep{telesco1991,knacke1993,okamoto2004,chen2007,li2012}. In general, it has been concluded that these are exceptional systems, in many cases possibly the sites of elevated dynamical activity that has temporarily boosted the amount of very small dust in their debris systems.

Here we present $\NumTargs$ additional warm debris disks with evidence for silicate emission features in their {\it Spitzer} Infrared Spectrograph (IRS; \citealp{houck2004}) data. The presence of a warm debris disk component was previously known to exist around these stars, but no analysis of their spectral features has been published\footnote{One exception is HIP86305. \citet{morales2013} created a model for this system that does reproduce the IRS emission feature, as we discuss in $\S$\ref{sec:specifictargets}}. The large number of new detections of features implies that such behavior is not exceptional. We fit to each spectrum physically-motivated SED models and determine the disk location and grain properties. We find that the locations of these disks can be well constrained, and that they are probing dust in the terrestrial regions of these systems. That is, analysis of subtle silicate features can be used to probe the terrestrial zones in many warm debris disks.       

\section{METHODS}
\label{sec:methods}

\subsection{Target Selection}
\label{sec:targetselection}

For stars identified in \citet{ballering2013} to host warm debris disk components, we inspected the IRS data for signs of spectral features. We limited our search to targets with data available from all four IRS low-resolution spectral orders (LL1, LL2, SL1, and SL2). We found $\NumTargs$ targets with signs of features out of the 106 warm components that had all four spectral orders; \citet{ballering2013} identified 125 total targets with warm components. The stars in our sample range in spectral type from B9 to F7, and their properties are given in Table \ref{table:targetlist}.

We inferred $T_\star$ and $M_\star$ from the known spectral types according to the tabulated values (or interpolations between those values) from \citet{carrollandostlie2006}. $L_\star$ values were computed from the bolometric magnitudes, as $L_\star = 10^{-0.4(M_\text{bol}-4.74)}L_\sun$, where $M_\text{bol} = V - 5\log_{10}(D) + 5 - A_V + BC$ and $A_V = 1.15(V-K-(V-K)_0)$. $BC$ is the bolometric correction inferred from the spectral type according to \citet{carrollandostlie2006}, and $(V-K)_0$ is the intrinsic color inferred from the spectral type according to \citet{cox2000}. We computed $R_\star$ from $T_\star$ and $L_\star$ using the Stefan-Boltzmann Law. MIPS 24 $\micron$ flux density values were taken from \citet{ballering2013}.

We used stellar ages from \citet{ballering2013} when available. These were estimated by combining chromospheric activity, x-ray emission, placement on the HR diagram, surface gravity, membership in clusters and associations, and gyrochronology. The references for these measurements are given in Table \ref{table:targetlist}. We also provide a quality flag for the age accuracy, giving the number of independent age measurements with good agreement. When ages were not available from \citet{ballering2013}, we found age references in the literature from studies that used reliable HR diagram fitting \citep{nielsen2013,zorec2012,chen2014}. For these targets, the age uncertainty can be large, sometimes $\gtrsim 50 \%$. \citet{zorec2012} provided ages in terms of the fraction of the main sequence lifetime; we obtained total main sequence lifetimes for these targets from Table 45 of \citet{schaller1992}, the $M_\star$ versus main sequence lifetime relation for the stellar evolution models employed by \citet{zorec2012}. We used $M_\star$ values for our targets from \citet{zorec2012} when using this table.

\subsection{IRS Data Reduction}
\label{sec:irsdatareduction}

The IRS Astronomical Observation Requests (AORs) for our targets are listed in Table \ref{table:targetlist}. The basic reduction was performed using the Spectroscopic Modeling Analysis and Reduction Tool (SMART) software package \citep{higdon2004}, as detailed in \citet{ballering2013}. In summary, bad pixels were removed using IRSCLEAN, multiple Data Collection Events (DCEs) for each nod position were combined, the background was removed from each 2D spectrum by subtraction of the opposite nod, the 2D spectra were converted into 1D spectra using optimal 2 nod extraction \citep{lebouteiller2010}, and the 1D spectra from both nods were combined. The result was a wavelength, flux, and uncertainty vector of each spectral order for each AOR. The ``bonus" third order data were not used. We compared our results with the reduction provided by the Cornell AtlaS of Spitzer/IRS Sources\footnote{The Cornell Atlas of Spitzer/IRS Sources (CASSIS) is a product of the Infrared Science Center at Cornell University, supported by NASA and JPL. http://cassis.astro.cornell.edu} (CASSIS; \citealp{lebouteiller2011}) to check that any structures in our spectra -- potential spectral features -- were not unique to our reduction procedure. We found no serious discrepancies between the two reductions, although there was typically a systematic difference in the absolute flux level.

We used the MATLAB software package for subsequent data reduction and analysis. We examined each spectral order and trimmed data from the ends, where the data are less reliable. The exact location of the trimming was determined individually for each target by eye, although we ensured some degree of overlap remained between adjacent orders. Next, we cut outlying data points from each order. To do this, we fit each spectral order with a third-degree polynomial, calculated the standard deviation of the residuals around this fit, and then discarded points lying more than three standard deviations from the fit. In practice, this process removed very few data points (primarily large outliers). We avoided a more aggressive cutting procedure as we did not want to erase any signs of emission features from our data. We corrected flux offsets between the orders by applying a multiplicative correction factor to the LL1, SL1, and SL2 flux values to bring them in line with the LL2 data and with each other. These corrections were determined by eye, and were typically less than 5\% and almost always less than 10\%. We combined the data from the four orders by interleaving the data at the overlapping regions and then smoothing the entire spectrum by binning to a wavelength resolution of 0.03 $\micron$. The silicate features we found are in the middle of the SL1 and LL2 wavelength coverages so they are not significantly affected by the order matching procedure. We then normalized the entire IRS spectrum to agree with the measured Multiband Imaging Photometer for {\it Spitzer} (MIPS; \citealp{rieke2004}) flux at 24 $\micron$, as the absolute calibration of MIPS is known to be more accurate than that of IRS.

With reduced IRS data in hand, we used a variety of empirical fitting approaches that indicated the presence of features roughly at the positions expected for silicate emission. The features were generally too weak to detect when simply viewing the data by eye -- they only became evident after subtracting the contribution from the stellar photosphere. We set out to confirm the signs of features by fitting the spectra with physically-motivated disk models capable of reproducing the emission features. We present the details of our model fitting procedure in the following section. 

\subsection{Model Fitting}
\label{sec:modelsedfitting}

To explore the location of the dust grains producing the emission features, we carried out fits to the debris disk spectra, as discussed below. In summary, we found that fitting with a single dust belt almost always resulted in a ``belt" so broad that a better physical explanation would be two belts, and for 9 of the sources single belt models could not even produce acceptable fits. We therefore fit all the systems with two debris belts. These fits indicate that the features arise from fairly narrow rings, which, as we show in $\S$\ref{sec:terrestrialzones}, are largely confined to the terrestrial zones around these stars.

The observed flux from a single dust grain is given by
\begin{equation}
\label{eq:grainflux}
F_\nu(\lambda,a,T_\text{d}) = \left(\frac{a}{D}\right)^2 Q_\text{abs}(\lambda,a) \pi B_\nu(\lambda,T_\text{d}),
\end{equation}
where $D$ is the distance to the system, $a$ is the grain size, $T_\text{d}$ is the temperature of the dust grain, and $B_\nu$ is the blackbody function. $Q_\text{abs}(\lambda,a)$ is the efficiency at which a dust grain absorbs and emits light, which depends on the dust composition. We assumed that all dust grains were composed of amorphous olivine (MgFeSiO$_4$), and we obtained the optical constants $(n,k)$ for this material as a function of wavelength from 0.2 to 500 $\micron$ from \citet{dorschner1995}. We then used the Mie Theory code \texttt{miex} \citep{wolf2004} with these optical constants to compute $Q_\text{abs}(\lambda,a)$ for a range of grain sizes.

The dust temperature is set by the energy balance of absorbed and emitted stellar radiation,
\begin{align}
\label{eq:grainbalance}
& \int_0^\infty \left(\frac{R_\star}{r}\right)^2 \pi a^2 Q_\text{abs}(\lambda,a) \pi B_\lambda(\lambda,T_\star) \,\mathrm{d}\lambda \nonumber\\ &= \int_0^\infty 4\pi a^2 Q_\text{abs}(\lambda,a) \pi B_\lambda(\lambda,T_\text{d}) \,\mathrm{d}\lambda,
\end{align}
where $r$ is the distance between the dust grain and the star. This equation cannot be solved explicitly for $T_\text{d}$, however it can be solved for $r$:
\begin{equation}
\label{eq:graindistance}
r = \frac{R_\star}{2}\sqrt{\frac{\int_0^\infty Q_\text{abs}(\lambda,a) B_\lambda(\lambda,T_\star) \,\mathrm{d}\lambda}{\int_0^\infty Q_\text{abs}(\lambda,a) B_\lambda(\lambda,T_\text{d}) \,\mathrm{d}\lambda}}.
\end{equation}
For each target in our sample, we computed $r$ over a grid of input $a$ and $T_\text{d}$ values. We then inverted the tabulated results to find $T_\text{d}(a,r)$. The integrals in Equation \ref{eq:graindistance} were carried out using MATLAB's \texttt{trapz} function with 200 wavelength values logarithmically spaced between 0.2 and 500 $\micron$. When generating models we checked that no dust grains reached temperatures above 1550 K, olivine's sublimation temperature.

We assumed the dust grains were distributed in a ring between $r_\text{in}$ and $r_\text{out}$ with surface density $\Sigma(r) \propto r^{-q}$. We also assumed that the grain size distribution was $n(a) \propto a^{-p}$ from $a_\text{min}$ to $a_\text{max}$, and was identical at all $r$. The total flux from a debris belt is then
\begin{align}
\label{eq:beltflux}
F_{\nu,\text{belt}}(\lambda) = &A \int_{a_\text{min}}^{a_\text{max}} \int_{r_\text{in}}^{r_\text{out}} \left(\frac{a}{D}\right)^2 Q_\text{abs}(\lambda,a) \nonumber\\ &\times \pi B_\nu(\lambda,T_\text{d}) (2 \pi r) r^{-q} a^{-p} \,\mathrm{d}r \, \mathrm{d}a.
\end{align}
When generating models, we evaluated Equation \ref{eq:beltflux} by summing the integrand in 500 $\times$ 500 bins in $r$ and $a$, distributed logarithmically between the maximum and minimum values. The normalization $A$ was set so that the total mass of dust represented by the model was $10^{-10} M_\sun$. We tested our procedure by comparing our models with those generated by the Debris Disk Radiative Transfer Simulator\footnote{http://www1.astrophysik.uni-kiel.de/dds/index.html} (DDS; \citealp{wolf2005}). We found a very good agreement between the resulting theoretical spectra.

Because we had SL IRS data at wavelengths as short as $\sim$5 $\micron$, and the infrared excess generally arose at somewhat longer wavelengths, our spectra provided sufficient information to determine the brightness of the stellar photosphere without relying on photometry from other instruments at shorter wavelengths. Thus, we modelled the photosphere and the excess together, and the normalization of the photosphere was included as a free parameter in our fits. The stellar photosphere was assumed to emit as a blackbody of temperature $T_\star$. A blackbody is appropriate because the stars in our sample, mostly A and F types, have virtually no spectral features in the mid-IR.

First, we attempted to fit our data with the stellar photosphere plus a single belt model. We limited the number of free parameters by fixing the grain size distribution exponent to $p=3.65$, as suggested by \citet{gaspar2012}. We also fixed the maximum grain size to $a_\text{max}=1000$ $\micron$, as the largest grains in a disk generally contribute little to the total flux at these wavelengths, making this parameter difficult to constrain. We limited $q$ to vary from 0 to 2. Thus, the form of our model was
\begin{align}
\label{eq:onemodel}
F_{\nu,\text{model}}(\lambda) &= C_pB_\nu(\lambda,T_\star) \nonumber\\ &+  C_0F_{\nu,\text{belt}}(\lambda,r_\text{in},r_\text{out},q,a_\text{min}),
\end{align}
with six free parameters $r_\text{in}$, $r_\text{out}$, $q$, $a_\text{min}$, $C_p$, and $C_0$.

Our fitting procedure entailed first defining a broad 4D grid of belt parameters and generating the model spectra for all points in this grid. We then found the best fit to the data (the optimal $C_p$ and $C_0$) for each model using MATLAB's \texttt{lsqcurvefit} algorithm by minimizing the standard $\chi^2$ metric (calculated in linear space). The parameter set corresponding to the lowest overall $\chi^2$ was deemed the best model. We examined how the minimum $\chi^2$ varied with each disk parameter to determine if we had located a global minimum in parameter space. We then created a revised parameter grid with greater precision centered on the previous minimum and repeated the fitting procedure, iterating this process until we were confident that the overall best fit was found. We found acceptable one-belt fits to $\NumOneBelt$ of the $\NumTargs$ targets. The parameters of the best fits for these targets are given in Table \ref{table:1beltresults}, and the the fits are shown in Figure \ref{fig:onebeltfits}. 

The IRS SL1 order can show spurious excess signal at 13.5 to 15 $\micron$ due to the ``SL 14 micron teardrop" effect (IRS Instrument Handbook). This artifact is thought to be caused by an internal reflection in the instrument, and is evident as a teardrop shape overlapping and slightly to the left of the spectral trace on the detector. We examined the 2D spectra of several sources and noticed signs of this effect. To avoid mistaking the teardrop signal for an emission feature, we excluded the 13.5 to 15 $\micron$ data for all of our targets when fitting models. However, in general the models fit this spectral range reasonably well (see Figures \ref{fig:onebeltfits} and \ref{fig:twobeltfits}) and including it in our $\chi^2$ minimization would not have modified the fits significantly.

\begin{figure*}
\centering
\includegraphics[scale=0.8]{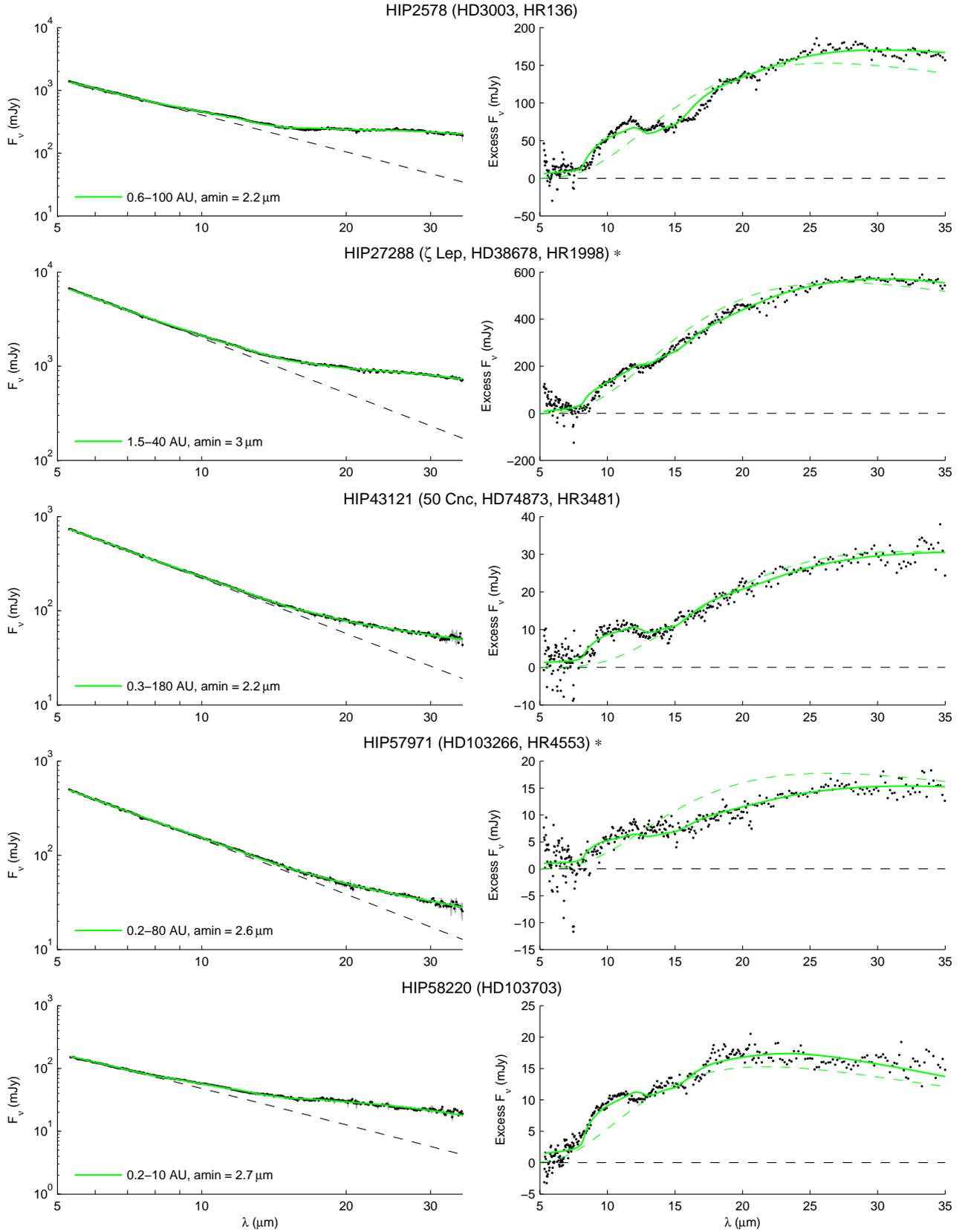}
\caption{Model spectra for the $\NumOneBelt$ targets that could be fit well by one belt. Targets marked with an asterisk have marginally detected features. IRS data are shown in black and the models are in solid green. The left panels show the total flux, while the right panels show the excess flux above the photosphere. Uncertainty in the data is shown in gray shading on the left panels, but is omitted from the right panels for clarity. Data between 13.5 and 15 $\micron$ were not included in the fitting procedure. The blackbody fits from \citet{ballering2013} are shown in dashed green (warm component) and dashed blue (cold component). Cases where the blackbody fits do not match the data well are due to the differences between this work and \citet{ballering2013} in how the stellar photosphere component was removed.}
\label{fig:onebeltfits}
\end{figure*}

\begin{figure*}
\centering
\figurenum{1}
\includegraphics[scale=0.8]{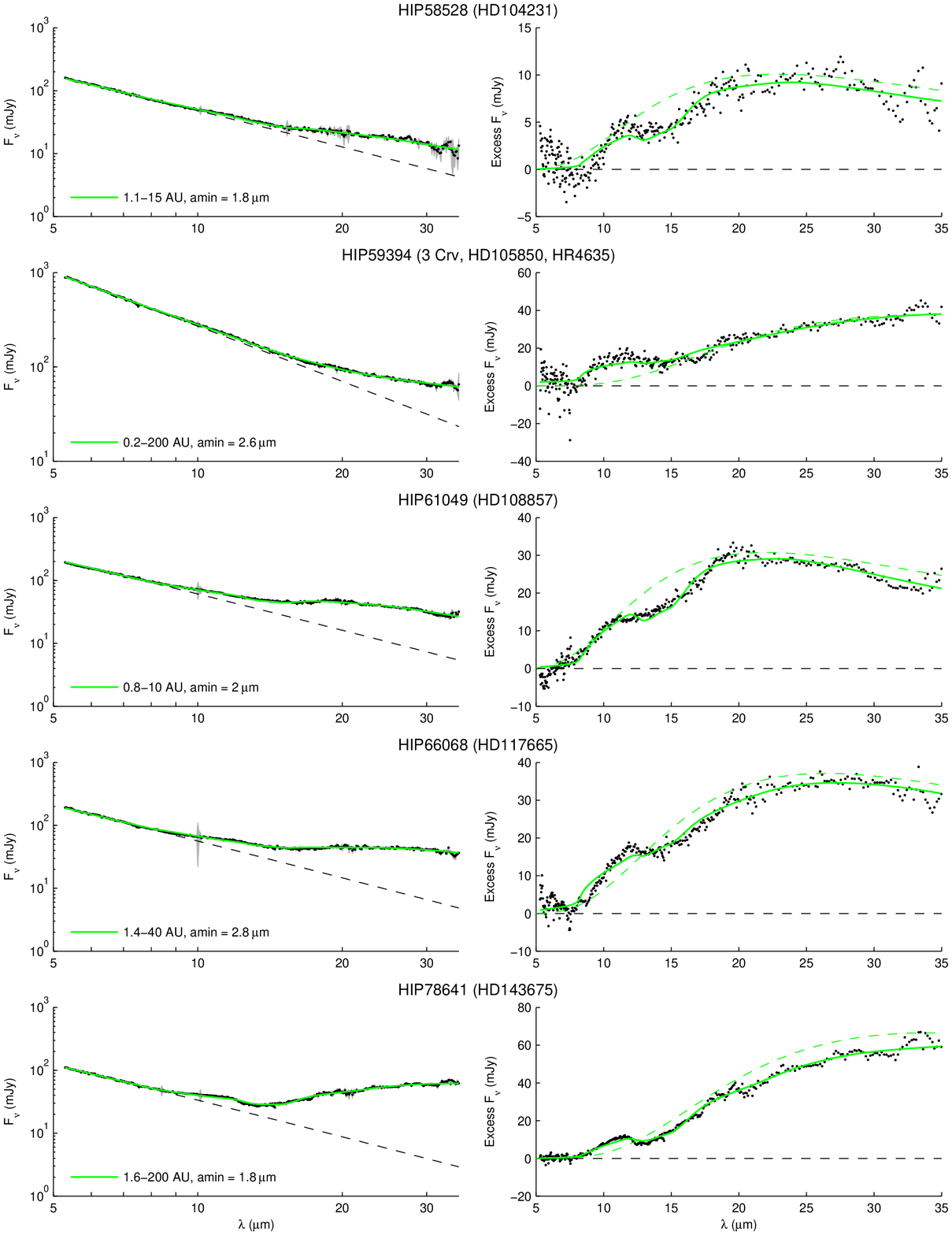}
\caption{(Continued)}
\end{figure*}

\begin{figure*}
\centering
\figurenum{1}
\includegraphics[scale=0.8]{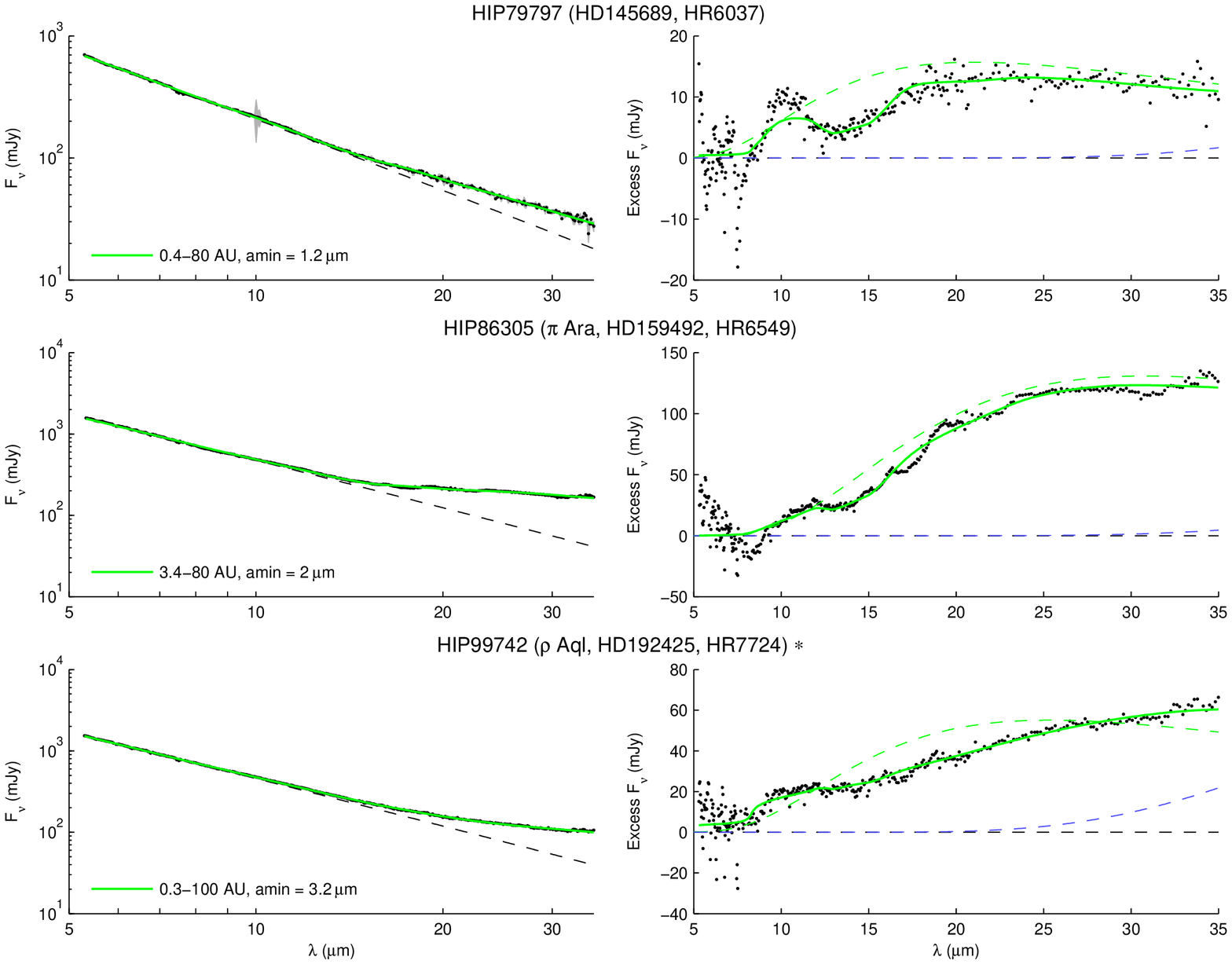}
\caption{(Continued)}
\end{figure*}

For the targets that could not be fit by one-belt models, the difficulty arose because the structure of the features at $\sim$10 $\micron$ and the levels of continuum excess at longer wavelengths could not both be reproduced with a single belt. Furthermore, the best fitting single-belt models were generally very large in radial extent, whereas many spatially resolved images of debris disks reveal them to be comprised of multiple, narrower belts. In fact, although all of the fits have inner radii within the terrestrial zones of the stars, 10 of the 13 fits have outer radii beyond 30 AU. Realistically, all of these fits could just as well be described as two-belt fits, since it is not plausible that there is a single component that is so broad. Therefore, we next fit all of our targets with two-belt models. We again fixed $p=3.65$ and $a_\text{max}=1000$ $\micron$ for both belts, and the form of the model was
\begin{align}
\label{eq:twomodel}
F_{\nu,\text{model}}(\lambda) &= C_pB_\nu(\lambda,T_\star) \nonumber\\ &+ C_1F_{\nu,\text{belt}}(\lambda,r_\text{in1},r_\text{out1},q_1,a_\text{min1}) \nonumber\\ &+ C_2F_{\nu,\text{belt}}(\lambda,r_\text{in2},r_\text{out2},q_2,a_\text{min2}),
\end{align}
with eleven free parameters $r_\text{in1}$, $r_\text{out1}$, $q_1$, $a_\text{min1}$, $r_\text{in2}$, $r_\text{out2}$, $q_2$, $a_\text{min2}$, $C_p$, $C_1$, and $C_2$. The fitting again entailed defining a 4D grid of parameters for the inner belt and for the outer belt and generating the single-belt model spectra for all points in these grids. We found the best fit to the data (the optimal $C_p$, $C_1$, and $C_2$) for each possible pairing of inner and outer models\footnote{As a point of clarity, when using these models (specified in terms of dust location), we refer to ``inner" and ``outer" belts. This is in contrast to blackbody models (specified in terms of dust temperature), for which we refer to ``warm" and ``cold" components.} such that $r_\text{out1} < r_\text{in2}$. The parameters of these best fits are presented in Table \ref{table:2beltresults}, and the fits are plotted in Figure \ref{fig:twobeltfits}. The fits were unable to constrain $q$; we generally found that we could fit the data nearly equally well while varying this parameter from 0 to 2 (although for the one-belt fits $q$ tended to be closer to 0).

For most targets, the 10 $\micron$ feature was fit almost entirely by the flux from the inner belt model; hence, the parameters of the inner belts were more constrained by our fitting than those of the outer belts, and we only report the parameters of the inner belts in Table \ref{table:2beltresults}. Virtually all the inner belts lie entirely within the terrestrial zones. 

In Tables \ref{table:1beltresults} and \ref{table:2beltresults} we give the total mass of dust (in grains from $a_\text{min}$ to 1000 $\micron$) for the belts, computed from $C_0 \times 10^{-10} M_\sun$ and $C_1 \times 10^{-10} M_\sun$. We also give the fractional luminosity of each belt, $L_\text{belt}/L_\star$, where the belt's emitting luminosity was calculated from
\begin{equation}
\label{eq:Lbelt}
L_\text{belt} = 4 \pi D^2 \int_{1\,\micron}^{2000\,\micron} \left(\frac{c}{\lambda^2}\right) F_{\nu,\text{belt}}(\lambda) \, \mathrm{d}\lambda.
\end{equation}

The uncertainties in our model fits were likely dominated by systematic errors, rather than by the statistical errors in the IRS flux density measurements. Calibration errors in the data were one source of systematic error, although we mitigated this by allowing the normalization of stellar photosphere flux density to be a free parameter in the fitting. Any errors in the stellar properties ($L_\star$, $T_\star$, D, etc.) influenced the models, adding systematic error to the best fit parameters. By using Mie theory we implicitly assumed the dust grains are spherical, but real grains are not perfect spheres, which added uncertainty to our models via our computed $Q_\text{abs}$ values. The robustness of our fits was also limited by using only one dust composition, fixing the form of $\Sigma(r)$ and $n(a)$, fixing $a_\text{max}$ and $p$, and using a maximum of two belts in our models. Varying the grain composition can result in changes in the radial scale of the belt to fit the observations, but variations by more than a factor of two are unlikely based on the range of optical constants available via the DDS website. Some silicate compositions fail to reproduce the shapes of the observed features entirely. The validity of our assumptions depends in part on the source of feature-producing dust, which we discuss in $\S$\ref{sec:discussion}. However, our best fit models generally reproduce the data well with physically reasonable parameters, so further increasing the model complexity and number of free parameters likely would have simply added degeneracies among the parameters. Furthermore, few if any of these additional free parameters could significantly undermine the detection of silicate features in the spectra.

\begin{figure*}
\centering
\includegraphics[scale=0.8]{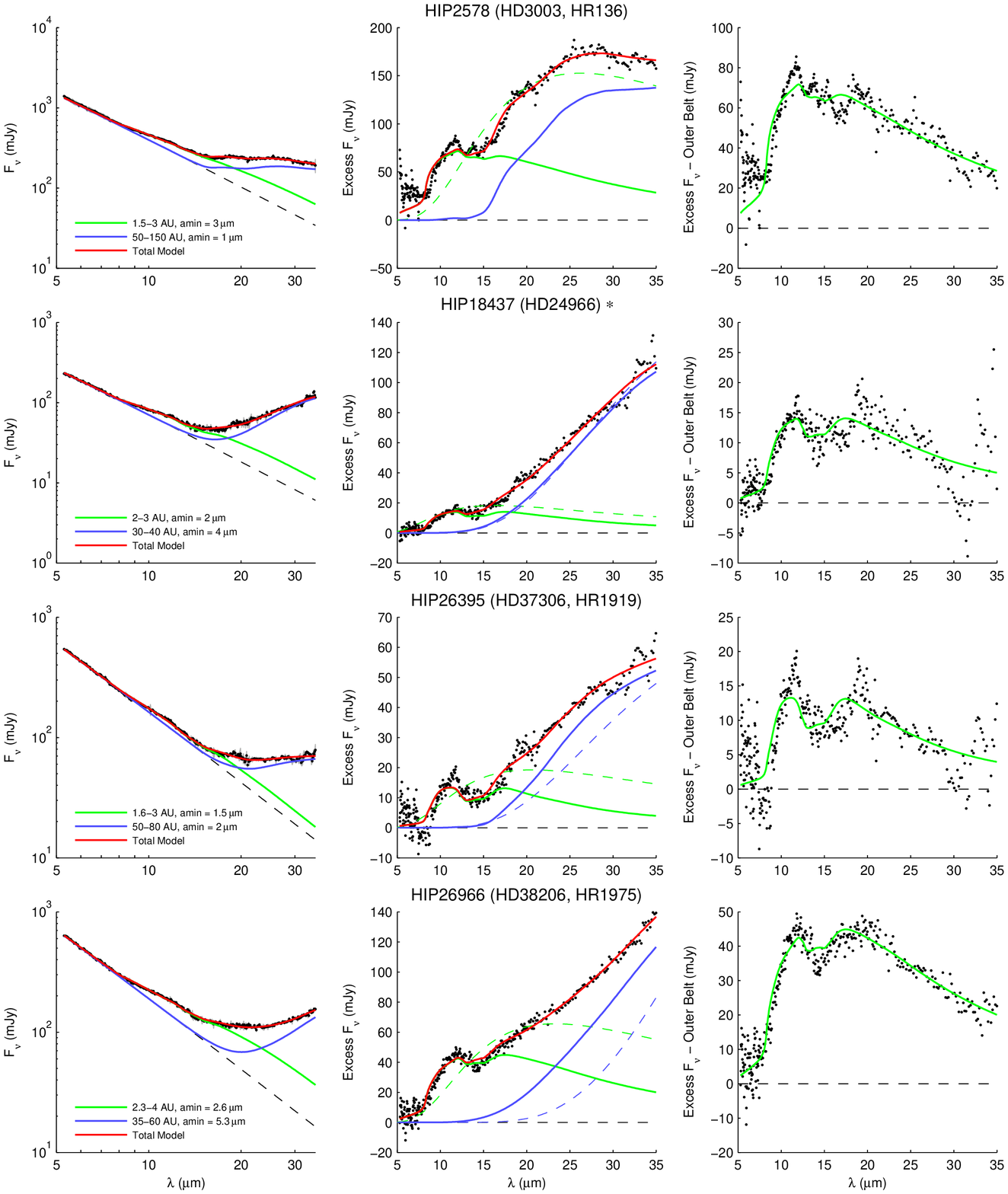}
\caption{Two-belt model fits for each target. Targets marked with an asterisk have marginally detected features. IRS data are shown in black, the inner belt models are in solid green, the outer belt models are in solid blue, and the total models are in solid red. The left panels show the total flux, the center panels show the excess flux above the photosphere, and the right panels show the remaining excess with the outer belt models removed.  Uncertainty in the data is shown in gray shading on the left panels, but is omitted from the other panels for clarity. Data between 13.5 and 15 $\micron$ were not included in the fitting procedure. The blackbody fits from \citet{ballering2013} are shown in the center panels in dashed green (warm component) and dashed blue (cold component). Cases where the blackbody fits do not match the data well are due to the differences between this work and \citet{ballering2013} in how the stellar photosphere component was removed.}
\label{fig:twobeltfits}
\end{figure*}

\begin{figure*}
\centering
\figurenum{2}
\includegraphics[scale=0.8]{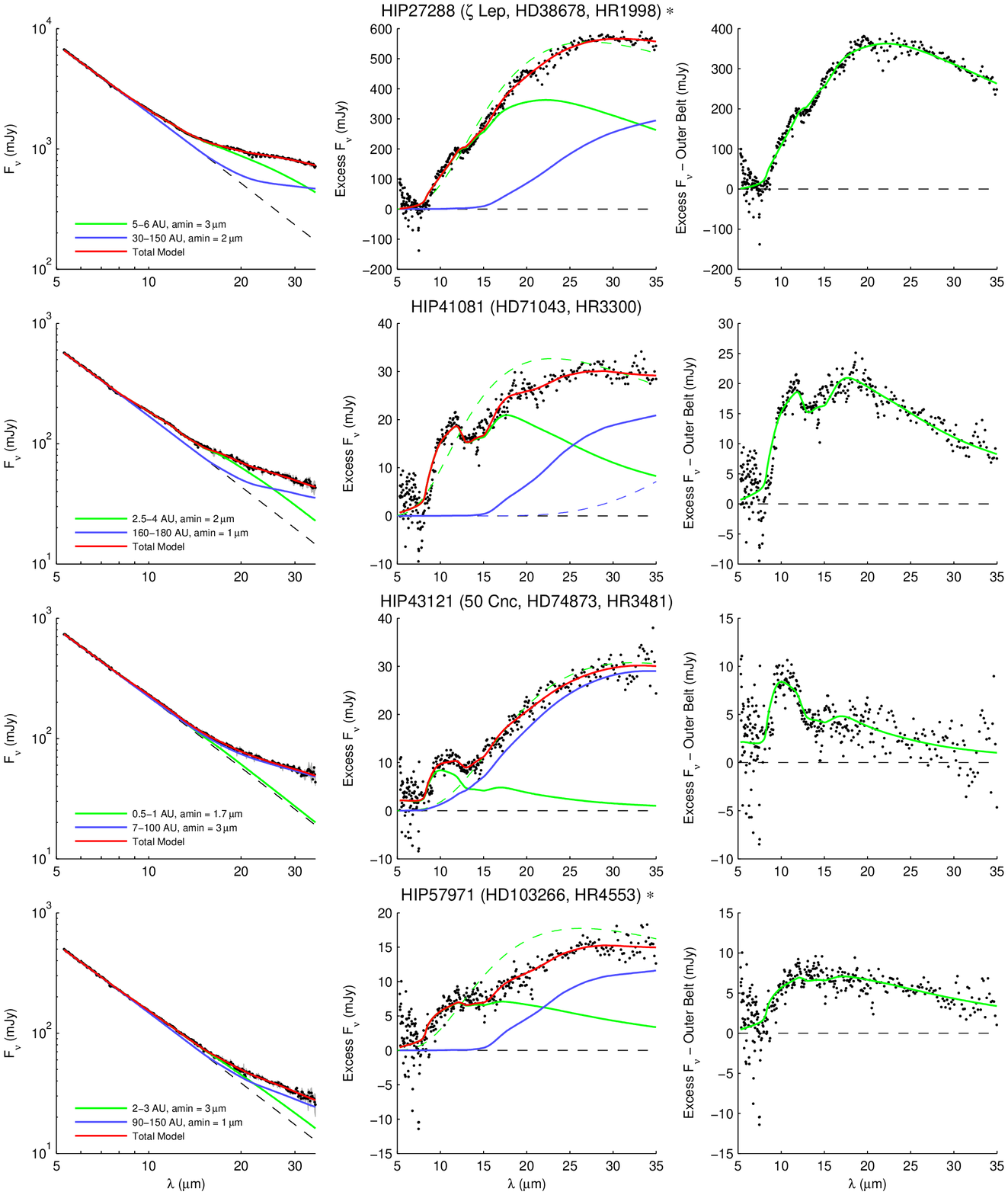}
\caption{(Continued)}
\end{figure*}

\begin{figure*}
\centering
\figurenum{2}
\includegraphics[scale=0.8]{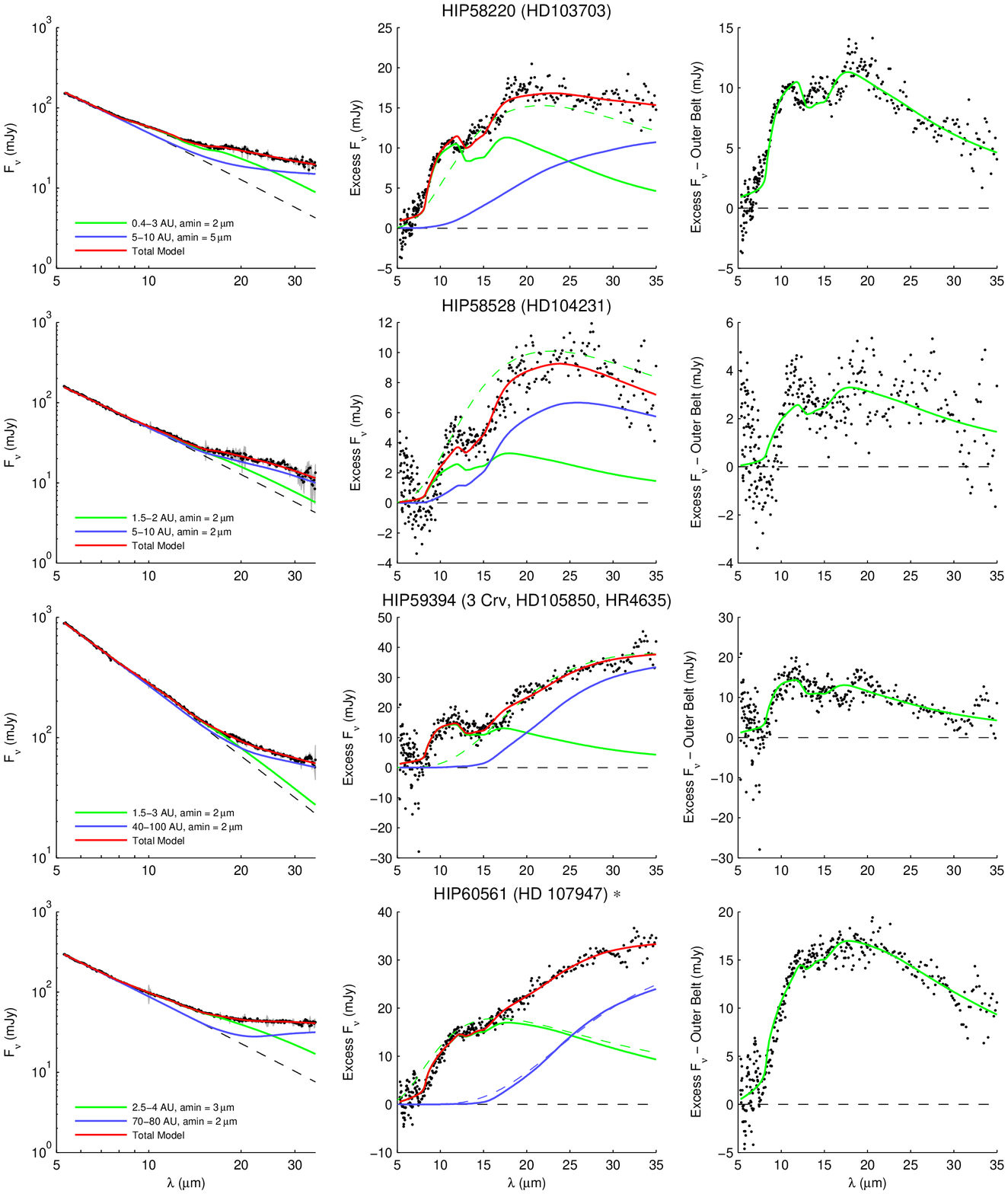}
\caption{(Continued)}
\end{figure*}

\begin{figure*}
\centering
\figurenum{2}
\includegraphics[scale=0.8]{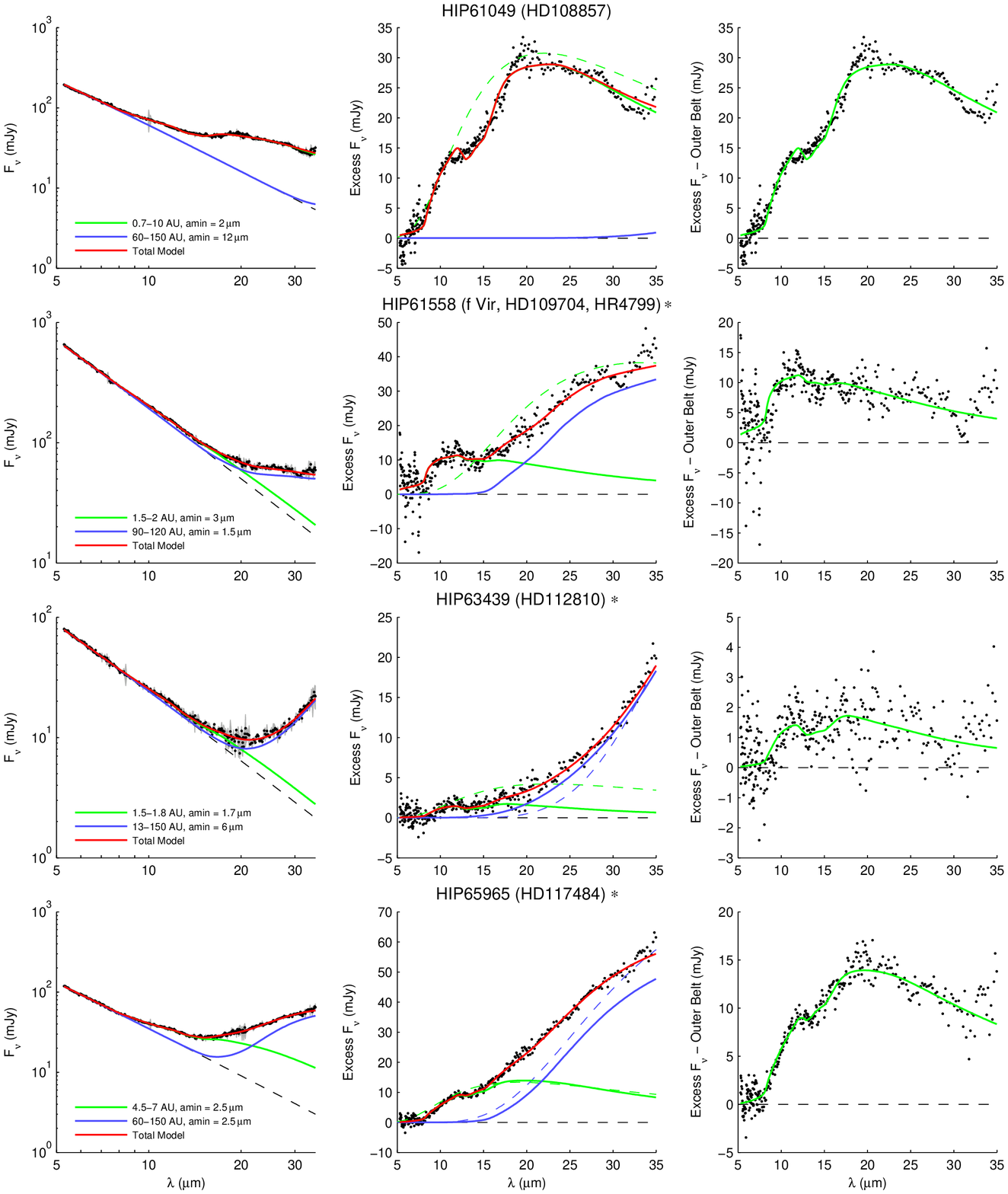}
\caption{(Continued)}
\end{figure*}

\begin{figure*}
\centering
\figurenum{2}
\includegraphics[scale=0.8]{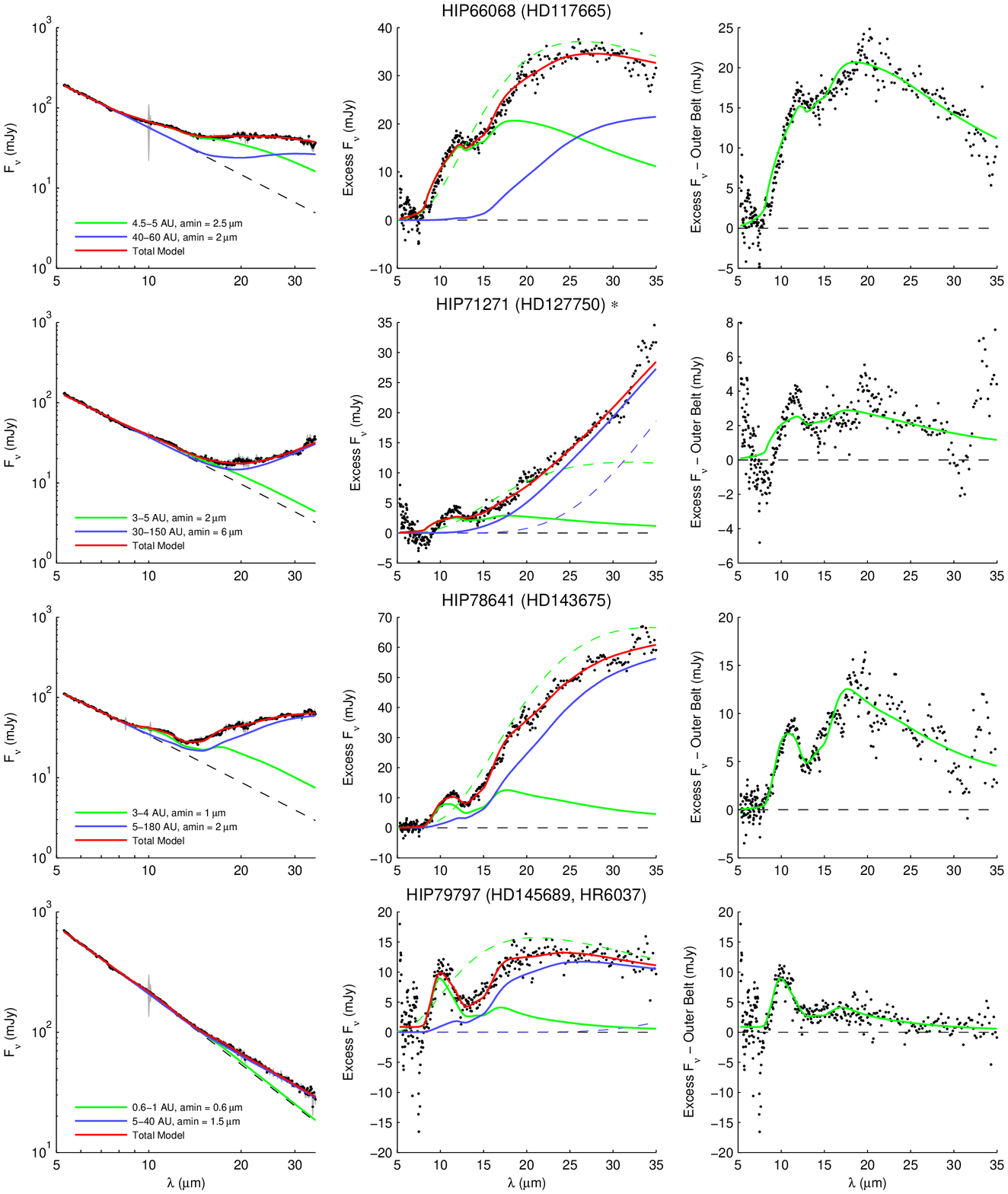}
\caption{(Continued)}
\end{figure*}

\begin{figure*}
\centering
\figurenum{2}
\includegraphics[scale=0.8]{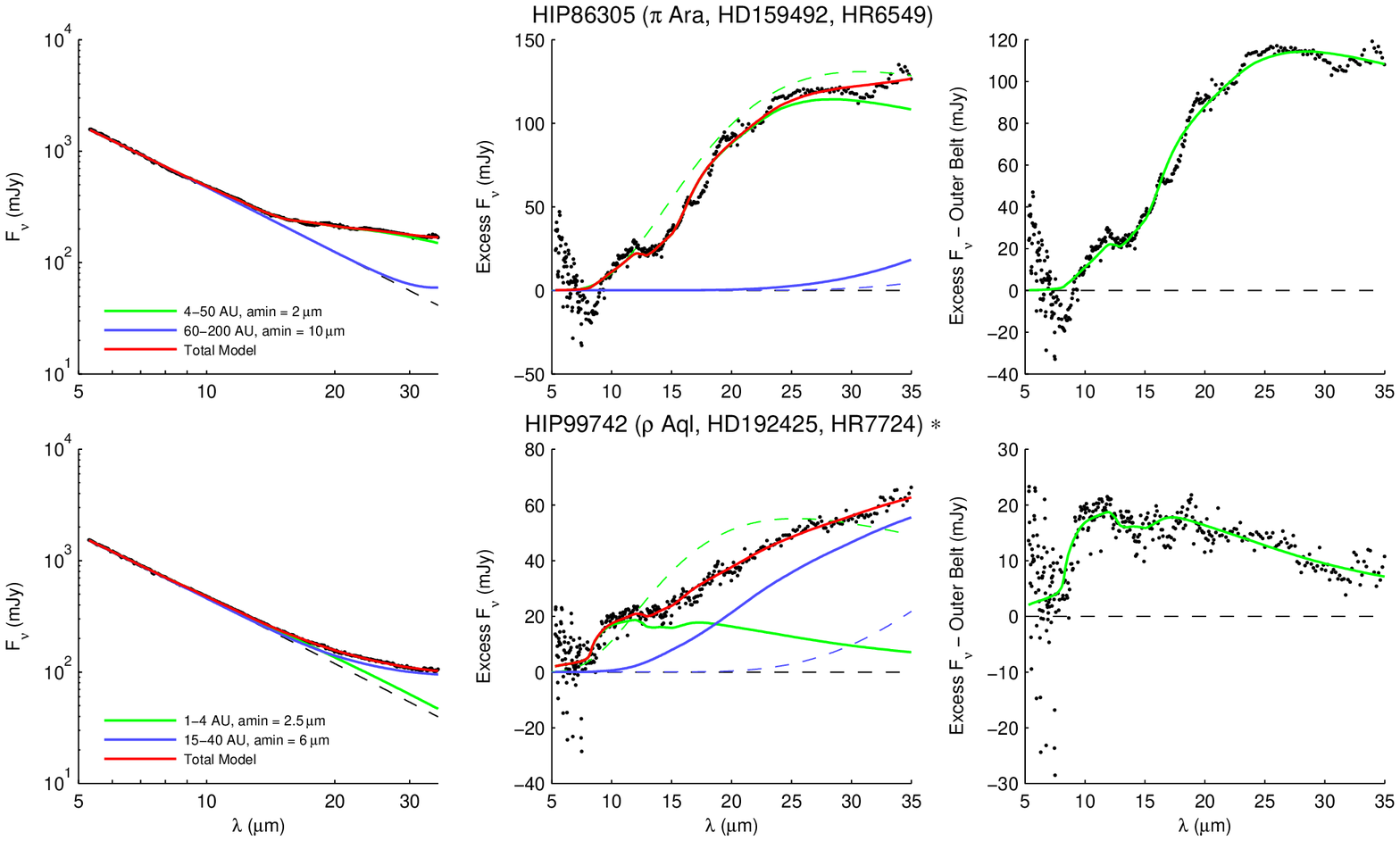}
\caption{(Continued)}
\end{figure*}

After inspecting the model fits, we segregated our targets into those with clear features and those with only marginally detected features. This designation is listed in Tables \ref{table:1beltresults} and \ref{table:2beltresults}, and targets with marginal features are marked with asterisks in Figures \ref{fig:onebeltfits} and \ref{fig:twobeltfits}. Some spectra with marginal features had only a very weak excess at 10 $\micron$, although the shape of the excess resembled a silicate emission feature. Others had a strong excess at 10 $\micron$, but its shape only differed slightly from a featureless blackbody. The majority of targets presented here do have clearly detected features, and these features vary in strength and the degree to which their signal is potentially confused by flux from the continuum. This suggests a natural variation in these features, predicting that there should be some low-level features that could be only marginally detected. Thus, it is likely that at least some of marginally-detected features here are true detections, and we include them in our target list for completeness.

\section{RESULTS}
\label{section:results}

\subsection{A Window to Terrestrial Zones}
\label{sec:terrestrialzones}

\begin{figure}
\centering
\includegraphics[scale=0.44,angle=0]{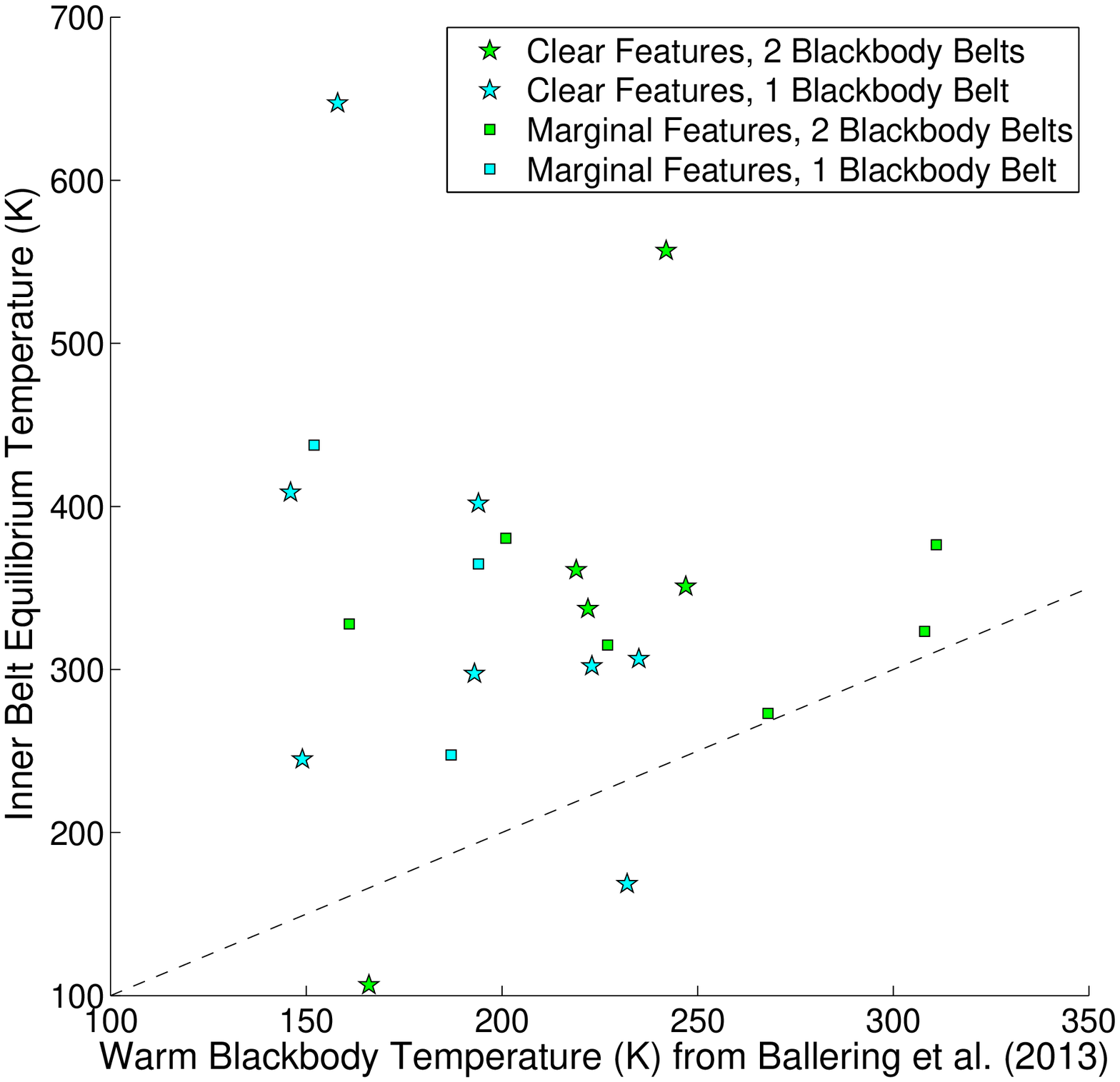}
\caption{The equilibrium temperatures of our best fitting models versus the temperatures of the warm component blackbody fits to these targets from \citet{ballering2013}. The equilibrium temperatures are calculated at the midpoint of the inner belt. Targets with clear features are stars and those with marginal features are squares. Green points are those that \citet{ballering2013} fit with two blackbodies and cyan points are those that \citep{ballering2013} fit with only one blackbody. This figure illustrates that fitting models to emission features can detect exozodiacal dust in the terrestrial zones of these systems (or even hotter zones), while the blackbody fits would find only asteroid belt zone dust.}
\label{fig:newvsoldtdust}
\end{figure}

Are the inner belts of these systems located in the terrestrial zones? The terrestrial zone is most easily defined in terms of an equilibrium temperature of $\sim$300 K. In Figure \ref{fig:newvsoldtdust} we plot the equilibrium temperatures of the dust in our best-fit models versus the temperatures of the warm components found by blackbody fitting from \citet{ballering2013} for these systems. We calculated the equilibrium temperatures using $T_\text{eq} = (\text{278.7 K})\left(\frac{L_\star}{L_\sun}\right)^{1/4} \left(\frac{r}{\text{1 AU}}\right)^{-1/2}$. Note that $T_\text{eq}$ is not the temperature of all dust grains in the belt at $r$, as the temperature varies significantly with grain size, and the smallest grains are significantly hotter than $T_\text{eq}$.

From the discussion in $\S$\ref{sec:modelsedfitting}, the single belt models are not plausible, either because they do not produce satisfactory fits (9 cases) or because they require belts that are so wide that they are in fact indicating the need for two belts. We computed $T_\text{eq}$ at the midpoint of the inner belt, $r = (r_\text{in1} + r_\text{out1})/2$. $T_\text{eq}$ values are listed in Table \ref{table:2beltresults}. We found that the inner belts are typically nearer to their stars than predicted by blackbody fitting. Within the errors, 18/22 of the inner belts lie within the terrestrial zones of their stars; there are two cases where the belts are too hot (HIP43121, HIP79797) and two where they are too cold (HIP61049, HIP86305).

It might seem surprising that disk models using realistic grain properties would predict dust belts to be nearer their stars than derived from blackbody fitting. As discussed in the Introduction, blackbody models tend to place debris disks closer to their stars than they actually are due to the presence of small, superheated grains. However, blackbody models often miss the emission features entirely (see the dashed lines in the right panels of Figure \ref{fig:onebeltfits} and the center panels of Figure \ref{fig:twobeltfits}). The signal of an emission feature (governed by $Q_\text{abs}$) is modulated by a blackbody function at the dust temperature (see Equation \ref{eq:grainflux}). Thus, disk spectra that show features are more likely to host an underlying population of dust at a temperature such that its blackbody peaks around 10 $\micron$. The spectral shape of the emission from this population of dust differs enough from a blackbody that a blackbody fitting routine is likely to ignore, rather than attempt to reproduce, the feature.

From our two-belt fits, we see that the flux from the feature-producing inner belt falls off quickly towards longer wavelengths, requiring an additional outer belt to fit the data. Because the flux from the inner belt is concentrated around the wavelength of the feature, the outer belt must account for more of the remaining flux than a cold component typically will when fitting with blackbodies. Indeed, the best fitting outer belt models for our targets are typically radially very broad. There may actually be three components in these systems, with dust in the terrestrial, asteroid belt, and Kuiper belt zones. Our inner belt models fit to the terrestrial zone dust, while our outer belt models accounted for all of the asteroid belt and the Kuiper belt zone dust that contributed to the IRS data. Our one-belt fits lead to a similar result: dust must be located at a wide range of radial locations, with some dust at least as close to the star as the terrestrial zone. The exact number of individual belts and the radial width of each belt cannot be determined from these data. However, the models require dust in the terrestrial zones to reproduce the emission features. Thus, the presence of these features is a useful tracer of exozodiacal dust in terrestrial zones.

\subsection{Notes on Specific Targets}
\label{sec:specifictargets}

Most of the debris disks in our sample are not well studied in the literature. The presence of a warm component was reported for most of these targets, but only from IRAS, MIPS, or WISE photometry. The IRS data for many of these targets were not published prior to \citet{ballering2013}. Some targets, however, have been well-studied in the past. We discuss these here, and how our discovery of emission features in their spectra fits with what was known about these systems. We also discuss notable aspects of our model fits for certain targets.

\textit{HIP2578 (HD3003, HR136) --} An infrared excess has been known to exist around HIP2578 for some time. The excess was detected with IRAS at 25 $\micron$ \citep{oudmaijer1992}, with MIPS at 24 and 70 $\micron$ \citep{smith2006}, and with AKARI at 18 $\micron$ but not at 9 $\micron$ \citep{fujiwara2013}. \citet{smith2010} detected the excess at 18 $\micron$ with ground based observations, but could not confirm an excess at 10 or 12 $\micron$. \citet{schutz2009} also detected no excess in an 8-13 $\micron$ spectrum obtained from the ground. At $\sim$10 $\micron$, these ground based measurements had uncertainties roughly as large or larger than the excess flux level we found from the IRS data; thus, the ground based observations are consistent with our results. The IRS spectrum was previously published by \citet{zuckerman2011} and by \citet{donaldson2012}. \citet{donaldson2012} fit these data and {\it Herschel} PACS photometry for this target with a single broad belt extending from 7.8 to 120 AU and a minimum grain size of 3.5 $\micron$. This fit also required an unusually steep grain size distribution with $p$=4.4. They noted no signs of emission features in the IRS spectrum. \citet{ballering2013} fit the IRS and MIPS data with a single blackbody of temperature 194 K, while \citet{chen2014} fit these data with two blackbodies at 472 and 173 K. Our inner belt location of 1.5-3 AU is consistent with the $<6.5$ AU constraint from (unresolved) images by \citet{smith2010}, although both our fits require an outer population of cold dust as well. HIP2578 may be a binary system of two A stars separated by 0\farcs1 (4.6 AU), measured in 1925 and 1964 \citep{dommanget1994,mason2001}. \citet{eggleton2008} report that HIP2578 is part of a 6 component system consisting of a wide hierarchical triple of three close binaries. The binarity of this system could impact both the heating and orbital stability of the debris disk. A detailed consideration of these factors is beyond the scope of this paper.

\textit{HIP26966 (HD38206, HR1975) --} The mid-IR excess around HIP26966 was detected by both IRAS \citep{mannings1998} and MIPS \citep{rieke2005,su2006}. The IRS data have been published several times \citep{morales2009,morales2011,zuckerman2011,ballering2013,chen2014}, but the presence of a 10 $\micron$ emission feature has not been previously discussed. \citet{smith2010} did not detect an excess at 10 $\micron$ with ground based observations. \citet{moerchen2010} also performed ground-based observations, detecting excess at 18.3 $\micron$ but not at 10.4 $\micron$. Their results are consistent with our measurements, but their uncertainties at 10.4 $\micron$ were too large to claim a significant excess detection. The 18.3 $\micron$ images of \citet{moerchen2010} were not spatially resolved, constraining the radius of the disk to $<$10.8 AU, which is consistent with our inner belt extending to 4 AU (nearly all of the excess flux at 18.3 $\micron$ is emanating from the inner belt in our model). No circumstellar dust was detected around this target in scattered light with HST \citep{krist2010}.

\textit{HIP27288 ($\zeta$ Lep, HD38678, HR1998) --} HIP27288 has long been known to host a warm debris disk. The mid-IR excess was detected by IRAS \citep{cote1987,aumann1991,mannings1998}, ISO \citep{habing2001}, MIPS \citep{rieke2005,su2006}, and a number of ground-based instruments \citep{fajardoacosta1998,chen2001,jayawardhana2001,schutz2005}. \citet{wyatt2007} noted that this excess is unusually bright for a system of its age, suggesting that this dust may be transient. No evidence for a very hot dust component was seen from near-IR interferometric observations \citep{absil2013}. \citet{chen2006} fit the IRS excess with a single 190 K blackbody, and noted no signs of emission features. This disk was also spatially resolved by ground-based observations at 18.3 $\micron$, locating the disk at 3 AU with some emission extending out to 8 AU \citep{moerchen2007}. Our inner belt (5-6 AU) is consistent with these results. Our models do differ significantly from the literature in that we include outer dust as well in the form of an outer belt in our two-belt model, or out to $\sim$40 AU in our one-belt model. There is clearly some structure in the IRS data at $\sim$10 $\micron$, but we classify the detection of a feature in this spectrum as marginal because this structure has a less peaked shape than features seen in other targets. If there is only one belt, then this structure must be the result of an emission feature. On the other hand, the structure could be the result of overlapping emission from multiple belts. A single blackbody model, however, is not a good fit to these data.

\textit{HIP43121 (50 Cnc, HD74873, HR3481) --} The debris disk around HIP43121 has not been particularly well studied in the literature. However, the IRS data were published by \citet{morales2009} and \citet{morales2011}. Both of these studies fit the excess with a single blackbody function at 190 K, but a footnote in \citet{morales2009} remarked that there were hints of a 10 $\micron$ feature in the data.

\textit{HIP58220 (HD103703) --} While fitting two-belt models to this target, we discovered a degeneracy in terms of which belt produced the majority of the 10 $\micron$ feature. Both local minima in parameter space produced fits with nearly identical $\chi^2$. The model we present in Figure \ref{fig:twobeltfits} relied on the inner belt to reproduce the feature while the outer belt resembled a nearly-smooth continuum, as was the case for most of our targets. In the alternative model, a broad outer belt with a small minimum grain size contributed significantly to the 10 $\micron$ feature while the inner belt with only larger grains provided primarily continuum. The parameters for this second case were: $a_\text{min1}$ = 6 $\micron$, $r_\text{in1}$ = 0.8 AU, $r_\text{out1}$ = 1 AU, $M_\text{dust1} = 0.218 \times 10^{-5} M_\earth$, $L_\text{belt1}/L_\star = 18.7 \times 10^{-5}$, $a_\text{min2}$ = 1 $\micron$, $r_\text{in2}$ = 1 AU, $r_\text{out2}$ = 100 AU, $M_\text{dust2}=73.7 \times 10^{-5} M_\earth$, and $L_\text{belt2}/L_\star=15 \times 10^{-5}$.

\textit{HIP61049 (HD108857) --} Like HIP58220, this target also exhibited a degeneracy in terms of which belt produced the emission feature. In Figure \ref{fig:twobeltfits} we show the fit where the inner belt could reproduce the entirety of the data with no need for an outer belt. In the other fit, the outer belt contributed significantly to the spectrum, including to the emission feature, due to having a smaller minimum grain size than the inner belt. The parameters of this fit were: $a_\text{min1}$ = 5 $\micron$, $r_\text{in1}$ = 0.7 AU, $r_\text{out1}$ = 0.8 AU, $M_\text{dust1}=0.17 \times 10^{-5} M_\earth$, $L_\text{belt1}/L_\star = 23.3 \times 10^{-5}, a_\text{min2}$ = 1.7 $\micron$, $r_\text{in2}$ = 4 AU, $r_\text{out2}$ = 7 AU, $M_\text{dust2} = 4.98 \times 10^{-5}$, and $L_\text{belt2}/L_\star = 26.1 \times 10^{-5}$.

\textit{HIP79797 (HD145689, HR6037) --} \citet{zuckerman2011} fit the MIPS and IRS excess with a 220 K blackbody, but did not mention any emission feature. The noteworthy aspect of this system is that the primary star is orbited at a projected separation of 350 AU by a binary system of brown dwarfs, separated from each other by 3 AU \citep{huelamo2010,nielsen2013}.

\begin{figure}
\centering
\includegraphics[scale=0.44,angle=0]{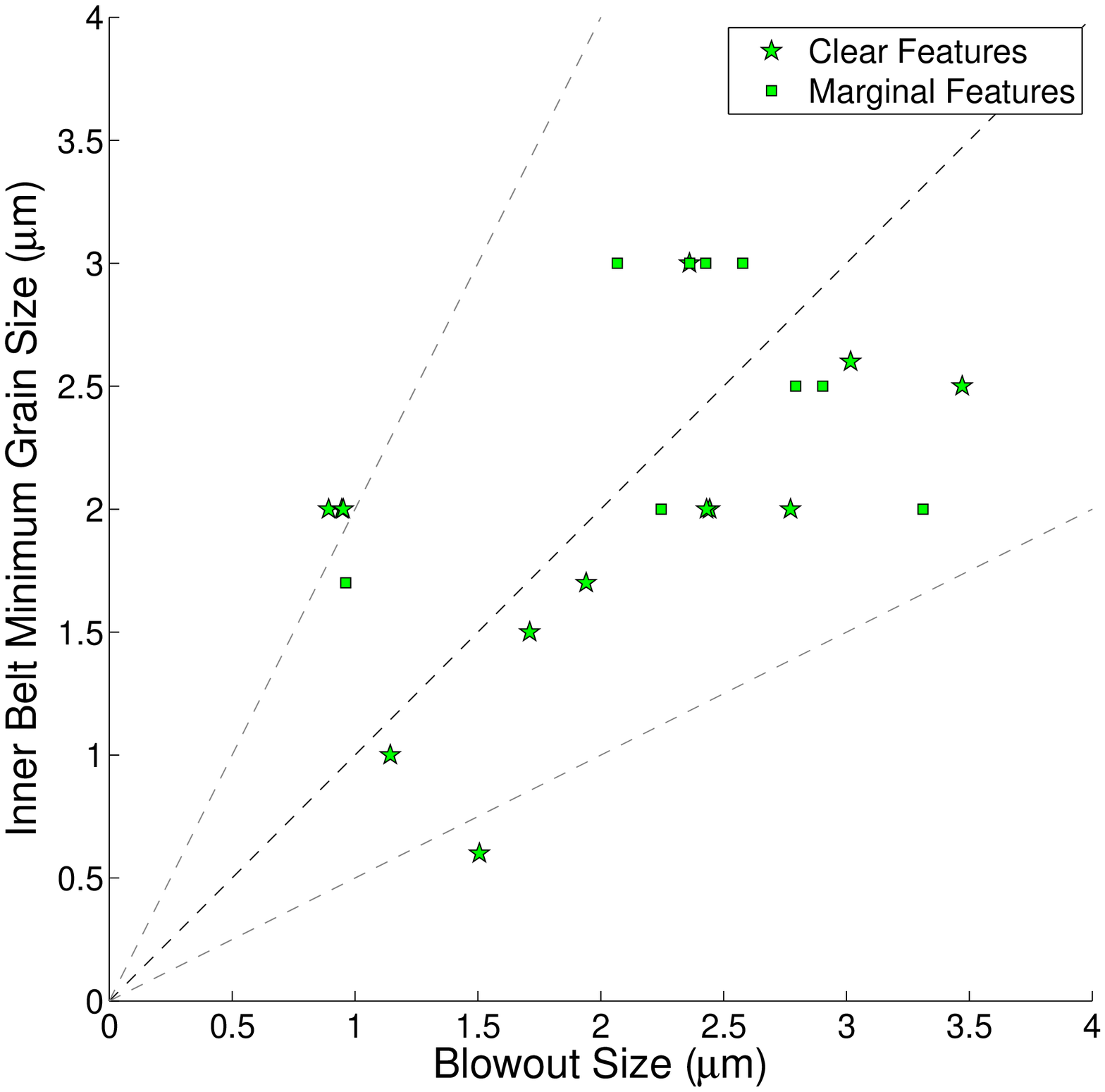}
\caption{This figure shows that $a_\text{min}$ is consistent with $a_\text{BOS}$ for our targets. Stars are targets with clear features and squares are those with marginal features. The bounding trend lines show the effect of varying $a_\text{BOS}$ by a factor of two (e.g., due to more complex grain structure). Low-level silicate emission features can arise in debris disk spectra, even when grains are not below the blowout size.}
\label{fig:aminvsabos}
\end{figure}

\textit{HIP86305 ($\pi$ Ara, HD159492, HR6549) --} Excess emission was first detected around HIP86305 with IRAS \citep{cheng1992,mannings1998}. No circumstellar dust was seen in scattered light with HST \citep{doering2007}. \citet{morales2009} and \citet{morales2011} fit blackbodies to the IRS excess and found a warm component temperature of 160 K, well outside of the terrestrial zone, although \citet{morales2009} noted that there may be signs of a faint spectral feature in the data. \citet{morales2013} resolved the outer edge of the outer belt with \textit{Herschel} (116 AU) and reanalyzed the IRS data, incorporating the constraint from \textit{Herschel}, and using physically-motivated belt models that could reproduce the structure in the IRS spectrum. They found an inner belt location of 9.1 or 9.8 AU, depending on the grain composition, which falls between the inner and outer edges of our best fit inner belt model. The analysis of \citet{morales2013} required grains three or four times smaller than the blowout size (requiring $a_\text{min}$ $\sim$1 $\micron$), whereas we found $a_\text{min1}$ consistent with the blowout size for this system (we found a larger minimum grain size and a slightly lower blowout size). \citet{morales2013} used a lower value for the stellar luminosity ($\sim$10 $L_\sun$) and different grain compositions, which may have contributed to the discrepancy with our results. Furthermore, neither we nor \citet{morales2013} could reproduce all of the structure seen in the IRS data, which may arise from crystalline grains.       

\begin{figure}
\centering
\includegraphics[scale=0.44]{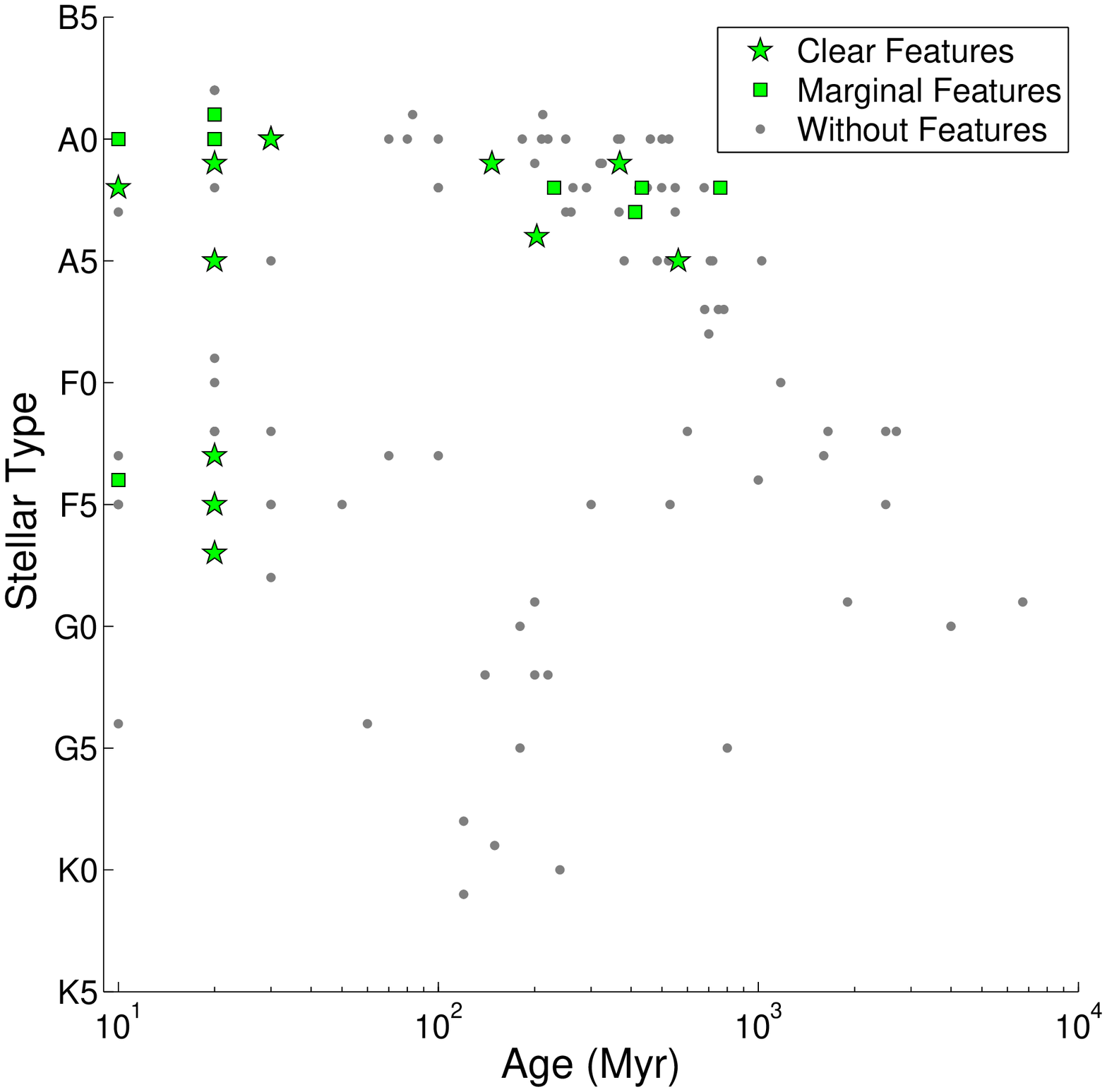}
\caption{The stellar type and age for targets with clear features (green stars), marginal features (green squares), and no features (gray circles). The properties of the targets with no features are from \citet{ballering2013}. This illustrates that features can be used to probe the terrestrial regions of young planetary systems over a range of stellar types, and also for older planetary systems of early type stars. Note that there are actually three 30 Myr-old A0 targets with clear features and two 20 Myr-old A0 targets with marginal features in our sample.}
\label{fig:typevsage}
\end{figure}

\section{DISCUSSION}
\label{sec:discussion}

Our analysis shows that a significant number of debris disks with excesses in the mid-IR exhibit low-level silicate emission features, indicating the presence of exozodiacal dust in their terrestrial zones. In this section we compare properties of these systems with others that host featureless warm debris disks, and we discuss potential sources for this exozodiacal dust.     

Dust produced by a steady state collisional cascade of planetesimals is expected to have a minimum grain size set by the blowout size of the system. Dust smaller than this size will be removed from the system by the star's radiation pressure. The presence of grains smaller than the blowout size could indicate that a large amount of dust was produced recently, such as in a massive collision. If the smallest grains were found to be significantly larger than the blowout size, then there might be other forces acting to remove small grains such as interactions with the ISM. We calculated the blowout size for each of our targets using Equation 5 of \citet{donaldson2012}:
\begin{align}
\label{eq:abos}
a_\text{BOS} = &(1.15 ~\micron)\left(\frac{L_\star}{L_\sun}\right) \left(\frac{M_\star}{M_\sun}\right)^{-1} \nonumber\\ &\times \left(\frac{\rho}{\text{1 g cm}^{-3}}\right)^{-1},
\end{align}
where $\rho$ is the density of the grain material (we assumed $\rho=3.71 \, \text{g cm}^{-3}$). The results for our targets are given in Tables \ref{table:1beltresults} and \ref{table:2beltresults}. In Figure \ref{fig:aminvsabos} we plot the minimum grain size of the inner belt of the two-belt models versus the blowout size for the targets. We see that the minimum grain size does track the blowout size, suggesting that these systems are not being influenced by rare or extreme circumstances. Equation \ref{eq:abos} assumes grains are solid spherical particles. Real grains may have more complex structures, so $a_\text{BOS}$ is likely only accurate to within a factor of $\sim$2.

\begin{figure}
\centering
\includegraphics[scale=0.44]{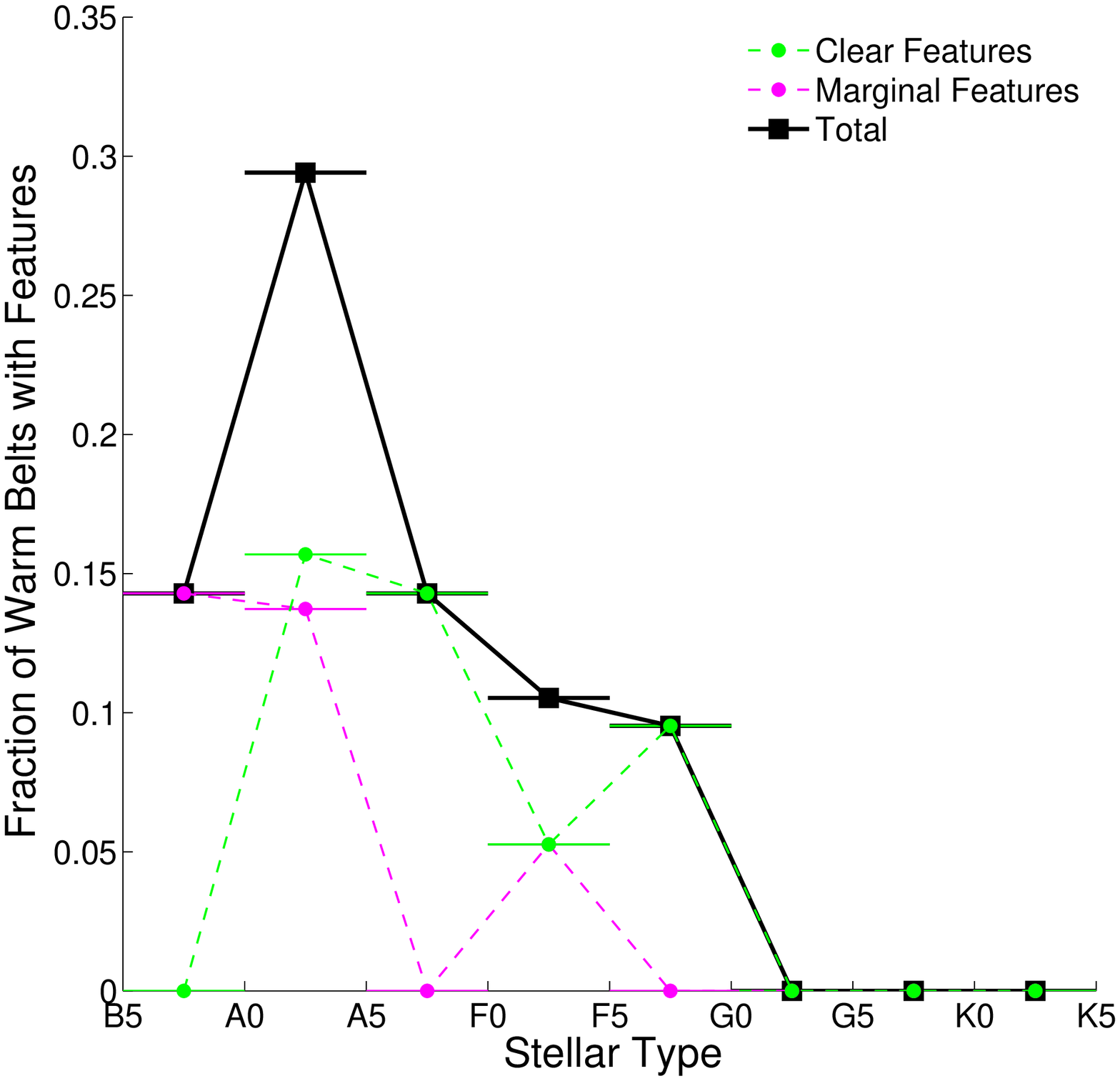}
\caption{The fraction of warm debris disks showing features in bins of stellar type. The fraction with clear features is in green, the fraction with marginal features is in magenta, and the total (the sum of green and magenta) is in black. Note that the fractions with clear and with marginal features are consistent with each other in their variation with stellar type.}
\label{fig:fracfeaturesvstype}
\end{figure}

\begin{figure}
\centering
\includegraphics[scale=0.44]{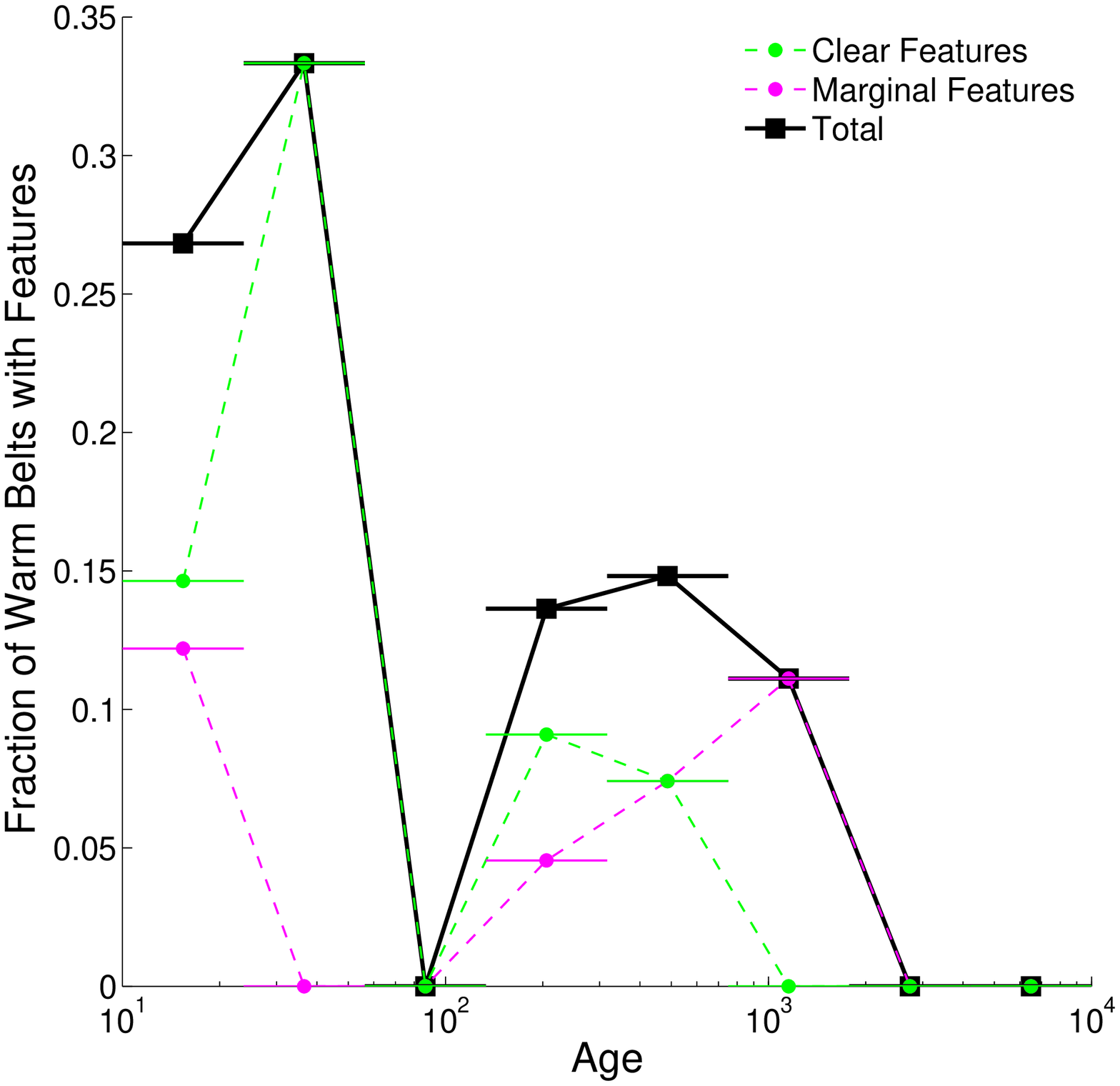}
\caption{The fraction of warm debris disks showing features in bins of system age. The fraction with clear features is in green, the fraction with marginal features is in magenta, and the total (the sum of green and magenta) is in black. Features are found around both young and older systems. Note that the fractions with clear and with marginal features are consistent with each other in their variation with age. The sharp spike at 25-60 Myr and minimum at 60-130 Myr may be a result of small number statistics, as there are few total warm disks in this age range.}
\label{fig:fracfeaturesvsage}
\end{figure}

We next investigated the age and stellar type of targets that exhibited features, as shown in Figure \ref{fig:typevsage}. The gray points are systems with warm disks from the sample of \citet{ballering2013} that did not show any features. For targets with no age determination in \citet{ballering2013}, we found age values from \citet{nielsen2013}, \citet{zorec2012}, and \citet{chen2014}, as we did for the targets with features (see $\S$\ref{sec:targetselection}). In Figures \ref{fig:fracfeaturesvstype} and \ref{fig:fracfeaturesvsage} we show the fraction of warm debris disks that exhibit features, binned by stellar type and by age.

We find that many of the disks with features are young ($\sim$10-30 Myr), but that there is a significant number of older disks (hundreds of Myr) that also have features. Young disks with features have stellar types spanning the range of the parent sample (late B through F), while the only older disks that show features are the early to mid A types. While the presence of features does not appear uniform across all stellar types and ages, it is clear that analyzing features in debris disks can provide a means to study the terrestrial zones of planetary systems with a large range of stellar types and ages. Disks with clear features and with marginal features are distributed in approximately the same way by stellar type and age, lending further evidence to the notion that there is natural variation in feature strengths for systems with dust in their terrestrial zones.

\begin{figure}
\centering
\includegraphics[scale=0.44]{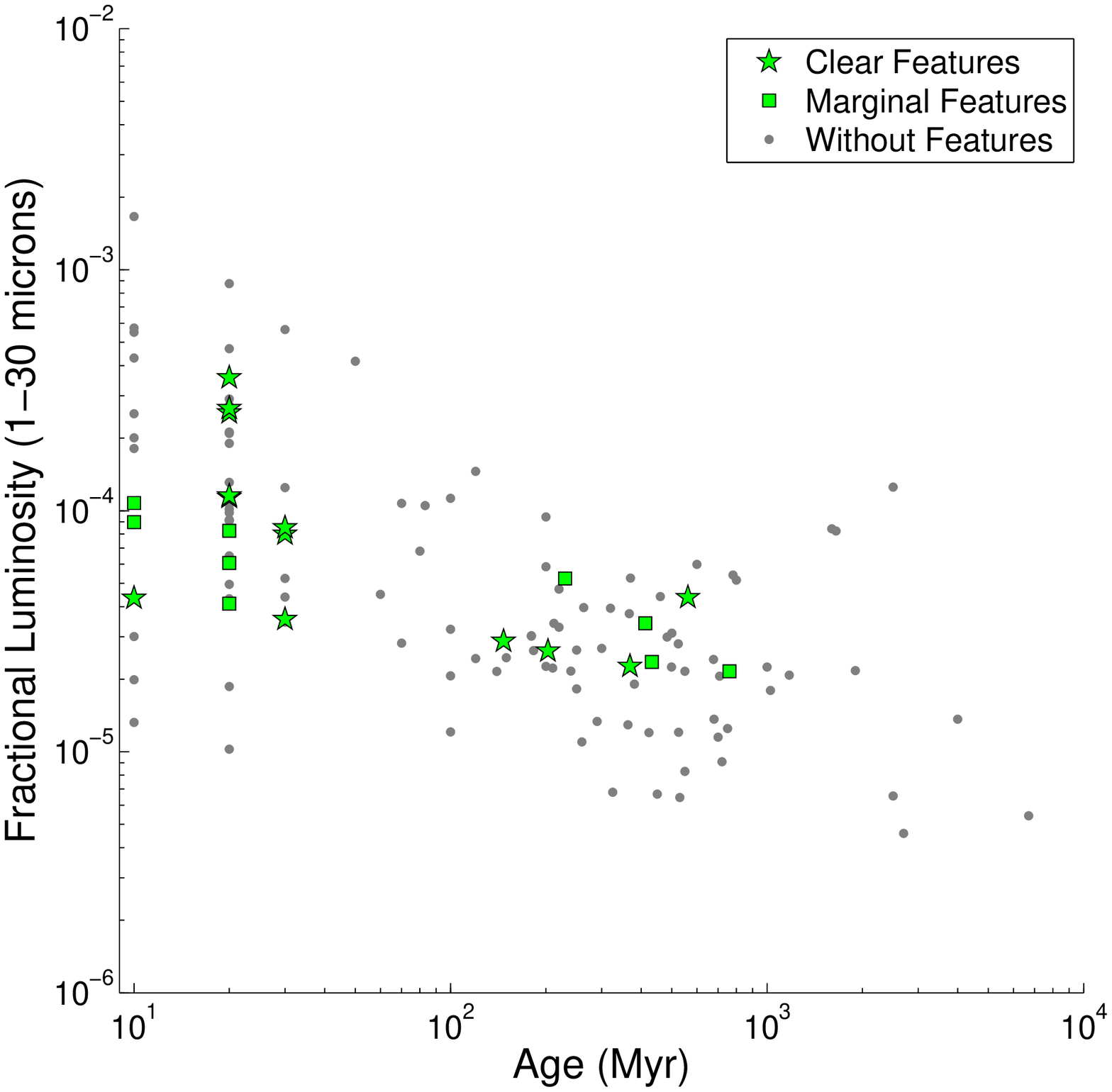}
\caption{The fractional luminosity (of the infrared excess from 1 to 30 $\micron$) and age of targets with clear features (green stars), marginal features (green squares), and no features (gray circles). The properties of the targets with no features are from \citet{ballering2013}. The overall decrease in disk brightness is well-known from previous debris disk studies. We see that the fractional luminosities of disks with features (clear or marginal) are consistent with those of featureless disks.}
\label{fig:fraclumvsage}
\end{figure}

Are features preferentially found in bright debris disks? In Figure \ref{fig:fraclumvsage} we plot the fractional luminosity of the mid-IR excess of these systems versus age. The gray points are again warm disks from \citet{ballering2013} that do not show signs of features. The fractional luminosity values in this plot are not the same values as given in Tables \ref{table:1beltresults} and \ref{table:2beltresults}. As discussed in $\S$\ref{sec:terrestrialzones} and illustrated in Figures \ref{fig:onebeltfits} and \ref{fig:twobeltfits}, the models we used in this paper fit the IRS data differently than did the one or two blackbody functions used by \citet{ballering2013}. We did not have acceptable one-belt fits for all targets in this paper, nor is it appropriate to compare the brightness of just our inner (outer) belt with the warm (cold) component from blackbody fitting, since the two fits in the two papers are constrained by different wavelength ranges. Comparing the total fractional luminosities of all components would also not be a valid comparison, as \citet{ballering2013} included MIPS 70 $\micron$ data in the fitting and so were sensitive to a significant amount of cold dust that was not measured in this study. To properly compare these two samples, we calculated the luminosity of the total model (inner + outer belts using our two-belt models; one or two blackbodies) using Equation \ref{eq:Lbelt}, but only over the wavelength range from 1 to 30 $\micron$.

Figure \ref{fig:fraclumvsage} shows that targets with features (either clear or marginal) are not extraordinary in terms of the brightness of their mid-IR excess. This suggests that the detection of features is not a selection effect limited to very bright disks. Some planetary systems have a detectable population of dust in their terrestrial zones while others have significantly less dust in this region, but otherwise the systems with features and without features are quite similar.  

What is the source of exozodiacal dust? There could be a belt of parent body planetesimals in the terrestrial zone undergoing a collisional cascade and producing the dust. This model can explain most standard debris disks with dust in the asteroid belt or Kuiper belt zones, although debris disks on smaller orbits are expected to grind down and dissipate on much shorter timescales. \citet{kennedy2013} investigated the possibility of in situ dust production to explain the detection of excesses at 12 $\micron$ with WISE photometry, which they interpreted as emission from exozodiacal dust. As in this study, they found that most systems with exozodiacal dust were young, but that some older systems had 12 $\micron$ excess as well. They could reproduce this finding by assuming that all systems start with a population of parent bodies in this zone that steadily collide and decay with time (this explains the young systems), but additionally there are occasional random collisions between the remaining parent bodies, as required to explain the exozodiacal dust in the older systems.

Another possibility is that the dust was produced farther out and migrated inwards via Poynting-Robertson (PR) drag or other drag forces. Theoretical studies find that the interaction of PR drag and dust sublimation can create a population of dust extending inwards from the parent body belt where the dust is created to the sublimation radius \citep{kobayashi2008,vanlieshout2014}.

Exozodiacal debris could also be delivered to terrestrial zones by planetesimals scattered inwards from an outer belt. \citet{bonsor2012} modelled this scenario and found that a chain of closely packed planets is required to move material inwards effectively. Furthermore, for older systems the outer belt must be located at a large orbital radius such that it can be massive enough to deliver sufficient material inwards without quickly grinding itself down. The inward scattering of planetesimals can occur at a higher rate and be sustained for a longer time period if the outermost planet is actively migrating outwards into the planetesimal belt \citep{bonsor2014}. The efficiency of this mechanism depends sensitively on the number, locations, and masses of the planets in the chain, and on the properties of the planetesimal belt. If this mechanism is the dominant source of the exozodiacal dust, the detection of silicate emission features also implies that these targets host rich planetary systems.

A dynamical instability in a planetary system can also deliver planetesimals to the terrestrial region. In this case, planets scatter each other, destabilizing their orbits and scattering many planetesimals. \citet{bonsor2013} simulated such events and found that their signals are short-lived (0.7-2.8 Myr). Detecting a significant number of systems that recently underwent a dynamical instability is unlikely, and this mechanism cannot be the dominant source of exozodiacal dust. 

Exozodiacal dust may be the byproduct of collisions associated with the final, chaotic stage of terrestrial planet formation. \citet{jackson2012} modeled the dust production from such a collision (analogous to our Moon-forming impact) and found that, after an initial spike in infrared excess from vapor condensates, a low but detectable amount of debris can persist in the terrestrial regions for at least 1 Myr and often longer than 10 Myr. Each terrestrial planet undergoes multiple giant impacts during the chaotic phase of its formation. Terrestrial planet formation is expected to have finished once the system reaches an age of 100 Myr, so this explanation is plausible for the younger targets in our sample, but not for the older systems.

Our targets potentially represent a different source of exozodiacal dust than the debris disks with previously studied features listed in $\S$\ref{sec:introduction}. The excesses in those systems tend to be anomalously bright, and the emission features are often very strong. Those systems may have very recently experienced massive collisions such that transient populations of dust (e.g. vapor condensates) are still present. These massive collisions may be part of the terrestrial planet formation process (for the younger systems) or from periods of dynamical instability.

\section{CONCLUSIONS}
\label{sec:conclusions}

In summary, we found low-level silicate emission features in the IRS spectra of $\NumTargs$ warm debris disks that were previously studied by \citet{ballering2013}. $\NumClear$ of these had clearly detected features, while $\NumMarginal$ were only marginally detected. We fit these data with physically-motivated models, allowing us to constrain the radial locations and minimum grain sizes of the dust belts more precisely than was possible when fitting with blackbodies. Our fits place dust in the terrestrial zones of these targets, which was missed when these data were fit with only blackbody functions. An outer population of dust was also required to fit the data. The minimum grain sizes of the terrestrial zone dust were consistent with the blowout sizes of these systems, and the mid-IR fractional luminosities of debris disks with features were comparable to those of warm debris disks without features, implying that disks with features are ``normal". The properties of systems with marginally detected features were distributed almost identically as those of systems with clear features, suggesting that many of the marginal cases are likely true emission features.

We found systems with features at a range of stellar types and ages (although no features were found around older, later type stars). The analysis of emission features in the spectra of unresolved debris disks provides a powerful method to probe the terrestrial zones of planetary systems at various stages of their evolution. These results will complement mid-IR interferometric studies of exozodiacal dust, allowing for the robust characterization of regions of planetary systems that have, until now, remained largely out of reach.

\acknowledgments
We would like to thank Kate Su for many helpful comments and suggestions. We also thank the referee for providing valuable feedback. We made use of the NASA/IPAC Infrared Science Archive, which is operated by the Jet Propulsion Laboratory, California Institute of Technology, under contract with NASA. This work is based on observations made with the {\it Spitzer} Space Telescope, which is operated by the Jet Propulsion Laboratory, California Institute of Technology under a contract with NASA.

\bibliographystyle{apj}

\clearpage
\begin{landscape}
\begin{deluxetable}{ccccccccccccccc}
\setlength{\tabcolsep}{0.001in} 
\tabletypesize{\scriptsize}
\tablewidth{0pt}
\tablecolumns{15}
\tablecaption{Target Properties \label{table:targetlist}}
\tablehead{\colhead{HIP} & \colhead{Other} & \colhead{Spectral} & \colhead{D} & \colhead{Age} & \colhead{Age\tablenotemark{a}} & \colhead{Age} & \colhead{$T_\star$} & \colhead{$L_\star$} & \colhead{$R_\star$} & \colhead{$M_\star$} & \colhead{V} & \colhead{K} & \colhead{$F_{\nu}$(24 $\micron$)} & \colhead{IRS} \\ \colhead{Identifier} & \colhead{Identifiers} & \colhead{Type} & \colhead{(pc)} & \colhead{(Myr)} & \colhead{Quality} & \colhead{References} & \colhead{(K)} & \colhead{($L_\sun$)} & \colhead{($R_\sun$)} & \colhead{$M_\sun$} & \colhead{(mag)} & \colhead{(mag)} & \colhead{(mJy)} & \colhead{AOR}}
\startdata
HIP2578 & HD3003, HR136 & A0V & 45.6 & 30 & 2 & 4,7 & 9800 & 22.09 & 1.64 & 2.90 & 5.07 & 4.99 & 232 $\pm$ 2.31 & 21789184 \\
HIP18437 & HD24966 & A0V & 105.8 & 10 & 1 & 2 & 9800 & 21.01 & 1.59 & 2.90 & 6.89 & 6.86 & 68.43 $\pm$ 0.67 & 21792256 \\
HIP26395 & HD37306, HR1919 & A2V & 62.9 & 10 & 2 & 4 & 9020 & 13.42 & 1.50 & 2.43 & 6.09 & 5.97 & 65.46 $\pm$ 0.64 & 21794560 \\
HIP26966 & HD38206, HR1975 & A0V & 75.1 & 30 & 1 & 2,3 & 9800 & 28.22 & 1.85 & 2.90 & 5.73 & 5.78 & 112 $\pm$ 1.1 & 12713472 \\
HIP27288 & $\zeta$ Lep, HD38678, HR1998 & A2IV-Vn: & 21.6 & 230 & 1 & 1 & 9020 & 19.04 & 1.79 & 2.43 & 3.54 & 3.29 & 878 $\pm$ 8.78 & 4932864 \\
HIP41081 & HD71043, HR3300 & A0V & 70 & 30 & 2 & 3 & 9800 & 22.86 & 1.66 & 2.90 & 5.89 & 5.87 & 59.79 $\pm$ 0.63 & 16177408 \\
HIP43121 & 50 Cnc, HD74873, HR3481 & A1V & 54 & 147 & 1 & 8 & 9400 & 16.53 & 1.54 & 2.64 & 5.89 & 5.55 & 66.88 $\pm$ 0.68 & 14140928 \\
HIP57971 & HD103266, HR4553 & A2V & 74.1 & 760 & 1 & 10 & 9020 & 18.53 & 1.77 & 2.43 & 6.17 & 5.99 & 41.49 $\pm$ 0.44 & 21800704 \\
HIP58220 & HD103703 & F3V & 98.9 & 20 & 2 & 5 & 6919 & 4.27 & 1.44 & 1.48 & 8.5 & 7.39 & 25.87 $\pm$ 0.3 & 16170496 \\
HIP58528 & HD104231 & F5V & 110.5 & 20 & 2 & 5 & 6650 & 4.27 & 1.56 & 1.40 & 8.54 & 7.42 & 18.6 $\pm$ 0.33 & 21800960 \\
HIP59394 & 3 Crv, HD105850, HR4635 & A1V & 58.8 & 369 & 1 & 9 & 9400 & 23.61 & 1.84 & 2.64 & 5.47 & 5.32 & 80.24 $\pm$ 0.83 & 21801216 \\
HIP60561 & HD107947 & A0V & 91.1 & 20 & 2 & 5 & 9800 & 19.34 & 1.53 & 2.90 & 6.6 & 6.62 & 43.46 $\pm$ 0.44 & 22800896 \\
HIP61049 & HD108857 & F7V & 97 & 20 & 2 & 5 & 6388 & 3.86 & 1.61 & 1.26 & 8.6 & 7.07 & 40.07 $\pm$ 0.44 & 22802432 \\
HIP61558 & f Vir, HD109704, HR4799 & A3V & 69 & 412 & 1 & 9 & 8710 & 18.81 & 1.91 & 2.26 & 5.88 & 5.7 & 62.69 $\pm$ 0.65 & 21801984 \\
HIP63439 & HD112810 & F4IV/V & 143.3 & 10 & 2 & 5 & 6784 & 4.49 & 1.54 & 1.45 & 9.14 & 8.04 & 10.33 $\pm$ 0.17 & 21802752 \\
HIP65965 & HD117484 & B9V & 147.3 & 20 & 2 & 5 & 10500 & 30.78 & 1.68 & 3.29 & 7.53 & 7.52 & 39.31 $\pm$ 0.41 & 22803200 \\
HIP66068 & HD117665 & A1/A2V & 147.9 & 20 & 2 & 5 & 9400 & 29.57 & 2.06 & 2.64 & 7.21 & 7.08 & 43.71 $\pm$ 0.46 & 22801920 \\
HIP71271 & HD127750 & A0V & 175.7 & 20 & 2 & 6 & 9800 & 30.98 & 1.94 & 2.90 & 7.6 & 7.54 & 19.51 $\pm$ 0.2 & 26312704 \\
HIP78641 & HD143675 & A5IV/V & 113.4 & 20 & 2 & 5 & 8190 & 7.38 & 1.35 & 2.00 & 8.04 & 7.62 & 51.64 $\pm$ 0.51 & 22806528 \\
HIP79797 & HD145689, HR6037 & A4V & 52.2 & 203 & 1 & 8 & 8434 & 10.30 & 1.51 & 2.12 & 5.95 & 5.66 & 52.24 $\pm$ 0.53 & 21809152 \\
HIP86305 & $\pi$ Ara, HD159492, HR6549 & A5IV-V & 44.6 & 562 & 1 & 10 & 8190 & 15.68 & 1.97 & 2.00 & 5.25 & 4.78 & 198.3 $\pm$ 1.97 & 14143232 \\
HIP99742 & $\rho$ Aql, HD192425, HR7724 & A2V & 46 & 433 & 1 & 9 & 9020 & 21.91 & 1.92 & 2.43 & 4.95 & 4.77 & 132.6 $\pm$ 1.32 & 14143744 \\
\enddata
\tablenotetext{a}{1 means there was a single age determination, 2 means there were two independent and consistent age determinations.}
\tablerefs{
(1) \citet{vican2012} -- isochrone ages;
(2) \citet{rhee2007};
(3) \citet{su2006};
(4) \citet{tetzlaff2010};
(5) \citet{rizzuto2011};
(6) \citet{hoogerwerf2000};
(7) \citet{zuckerman2004};
(8) \citet{nielsen2013};
(9) \citet{zorec2012};
(10) \citet{chen2014}.
}
\end{deluxetable}
\clearpage
\end{landscape}

\clearpage
\begin{deluxetable}{ccccccccc}
\tabletypesize{\scriptsize}
\tablewidth{0pt}
\tablecolumns{9}
\tablecaption{One Belt Fitting Results \label{table:1beltresults}}
\tablehead{\colhead{HIP} & \colhead{Other} & \colhead{$a_\text{BOS}$} & \colhead{$a_\text{min}$} & \colhead{$r_\text{in}$} & \colhead{$r_\text{out}$} & \colhead{$M_\text{dust}$} & \colhead{$L_\text{belt}/L_\star$} & \colhead{Feature} \\ \colhead{Identifier} & \colhead{Identifiers} & \colhead{($\micron$)} & \colhead{($\micron$)} & \colhead{(AU)} & \colhead{(AU)} & \colhead{($\times 10^{-5} M_\earth$)} & \colhead{($\times 10^{-5}$)} & \colhead{Detection}}
\startdata
HIP2578 & HD3003, HR136 & 2.4 & 2.2 & 0.6 & 100 & 83.5 & 10.5 & clear \\ 
HIP27288 & $\zeta$ Lep, HD38678, HR1998 & 2.4 & 3 & 1.5 & 40 & 19 & 7.94 & marginal \\ 
HIP43121 & 50 Cnc, HD74873, HR3481 & 1.9 & 2.2 & 0.3 & 180 & 72.9 & 3.58 & clear \\ 
HIP57971 & HD103266, HR4553 & 2.4 & 2.6 & 0.2 & 80 & 16.7 & 3.52 & marginal \\ 
HIP58220 & HD103703 & 0.9 & 2.7 & 0.2 & 10 & 3.78 & 33.9 & clear \\ 
HIP58528 & HD104231 & 0.9 & 1.8 & 1.1 & 15 & 4.16 & 14.4 & clear \\ 
HIP59394 & 3 Crv, HD105850, HR4635 & 2.8 & 2.6 & 0.2 & 200 & 102 & 3.93 & clear \\ 
HIP61049 & HD108857 & 1.0 & 2 & 0.8 & 10 & 5.99 & 42.4 & clear \\ 
HIP66068 & HD117665 & 3.5 & 2.8 & 1.4 & 40 & 36.5 & 16.1 & clear \\ 
HIP78641 & HD143675 & 1.1 & 1.8 & 1.6 & 200 & 1420 & 50.2 & clear \\ 
HIP79797 & HD145689, HR6037 & 1.5 & 1.2 & 0.4 & 80 & 7.49 & 2.38 & clear \\ 
HIP86305 & $\pi$ Ara, HD159492, HR6549 & 2.4 & 2 & 3.4 & 80 & 54.4 & 7.3 & clear \\ 
HIP99742 & $\rho$ Aql, HD192425, HR7724 & 2.8 & 3.2 & 0.3 & 100 & 34.8 & 3.97 & marginal \\ 
\enddata
\tablecomments{$a_\text{BOS}$ is the blowout size for grains in the system, calculated from Equation \ref{eq:abos}. $a_\text{min}$ is the minimum grain size of our best-fit model. $r_\text{in}$ and $r_\text{out}$ are the inner and outer orbital radii of our best-fit model, respectively. $M_\text{dust}$ is the total mass of dust (in grains from $a_\text{min}$ to 1000 $\micron$) of our best-fit model. $L_\text{belt}/L_\star$ is the fractional luminosity of our best fit model, with $L_\text{belt}$ calculated from Equation \ref{eq:Lbelt}. The final column notes whether the detection of features was clear or marginal, as discussed in $\S$\ref{sec:modelsedfitting}.}
\end{deluxetable}

\begin{deluxetable}{cccccccccc}
\tabletypesize{\scriptsize}
\tablewidth{0pt}
\tablecolumns{10}
\tablecaption{Two Belt Fitting Results: Inner Belt Properties \label{table:2beltresults}}
\tablehead{\colhead{HIP} & \colhead{Other} & \colhead{$a_\text{BOS}$} & \colhead{$a_\text{min1}$} & \colhead{$r_\text{in1}$} & \colhead{$r_\text{out1}$} & \colhead{$T_\text{eq}$} & \colhead{$M_\text{dust1}$} & \colhead{$L_\text{belt1}/L_\star$} & \colhead{Feature} \\ \colhead{Identifier} & \colhead{Identifiers} & \colhead{($\micron$)} & \colhead{($\micron$)} & \colhead{(AU)} & \colhead{(AU)} & \colhead{(K)} & \colhead{($\times 10^{-5} M_\earth$)} & \colhead{($\times 10^{-5}$)} & \colhead{Detection}}
\startdata
HIP2578 & HD3003, HR136 & 2.4 & 3 & 1.5 & 3 & 402 & 0.277 & 6.12 & clear \\ 
HIP18437 & HD24966 & 2.2 & 2 & 2 & 3 & 376 & 0.285 & 5.98 & marginal \\ 
HIP26395 & HD37306, HR1919 & 1.7 & 1.5 & 1.6 & 3 & 351 & 0.0812 & 2.91 & clear \\ 
HIP26966 & HD38206, HR1975 & 3.0 & 2.6 & 2.3 & 4 & 361 & 0.653 & 7.34 & clear \\ 
HIP27288 & $\zeta$ Lep, HD38678, HR1998 & 2.4 & 3 & 5 & 6 & 248 & 1.58 & 5.36 & marginal \\ 
HIP41081 & HD71043, HR3300 & 2.4 & 2 & 2.5 & 4 & 337 & 0.252 & 3.26 & clear \\ 
HIP43121 & 50 Cnc, HD74873, HR3481 & 1.9 & 1.7 & 0.5 & 1 & 647 & 0.00538 & 1.72 & clear \\ 
HIP57971 & HD103266, HR4553 & 2.4 & 3 & 2 & 3 & 365 & 0.104 & 1.81 & marginal \\ 
HIP58220 & HD103703 & 0.9 & 2 & 0.4 & 3 & 306 & 0.341 & 21.6 & clear \\ 
HIP58528 & HD104231 & 0.9 & 2 & 1.5 & 2 & 302 & 0.133 & 6.25 & clear \\ 
HIP59394 & 3 Crv, HD105850, HR4635 & 2.8 & 2 & 1.5 & 3 & 409 & 0.0638 & 1.72 & clear \\ 
HIP60561 & HD107947 & 2.1 & 3 & 2.5 & 4 & 323 & 0.565 & 5.59 & marginal \\ 
HIP61049 & HD108857 & 1.0 & 2 & 0.7 & 10 & 168 & 5.83 & 43.4 & clear \\ 
HIP61558 & f Vir, HD109704, HR4799 & 2.6 & 3 & 1.5 & 2 & 438 & 0.0772 & 2.6 & marginal \\ 
HIP63439 & HD112810 & 1.0 & 1.7 & 1.5 & 1.8 & 315 & 0.0914 & 5.2 & marginal \\ 
HIP65965 & HD117484 & 2.9 & 2.5 & 4.5 & 7 & 273 & 1.86 & 6.22 & marginal \\ 
HIP66068 & HD117665 & 3.5 & 2.5 & 4.5 & 5 & 297 & 2.06 & 10.3 & clear \\ 
HIP71271 & HD127750 & 3.3 & 2 & 3 & 5 & 328 & 0.23 & 2.07 & marginal \\ 
HIP78641 & HD143675 & 1.1 & 1 & 3 & 4 & 245 & 0.632 & 11.5 & clear \\ 
HIP79797 & HD145689, HR6037 & 1.5 & 0.6 & 0.6 & 1 & 557 & 0.00323 & 1.73 & clear \\ 
HIP86305 & $\pi$ Ara, HD159492, HR6549 & 2.4 & 2 & 4 & 50 & 106 & 24.3 & 6.71 & clear \\ 
HIP99742 & $\rho$ Aql, HD192425, HR7724 & 2.8 & 2.5 & 1 & 4 & 380 & 0.0779 & 1.65 & marginal \\ 
\enddata
\tablecomments{$a_\text{BOS}$ is the blowout size for grains in the system, calculated from Equation \ref{eq:abos}. $a_\text{min1}$ is the minimum grain size of our best-fit model's inner belt. $r_\text{in1}$ and $r_\text{out1}$ are the inner and outer orbital radii of our best-fit model's inner belt, respectively. $T_\text{eq}$ is the equilibrium temperature at the midpoint of our best-fit model's inner belt. $M_\text{dust1}$ is the total mass of dust (in grains from $a_\text{min}$ to 1000 $\micron$) of our best-fit model's inner belt. $L_\text{belt1}/L_\star$ is the fractional luminosity of our best fit model's inner belt, with $L_\text{belt}$ calculated from Equation \ref{eq:Lbelt}. The final column notes whether the detection of features was clear or marginal, as discussed in $\S$\ref{sec:modelsedfitting}.}
\end{deluxetable}
 
\end{document}